\documentclass[twocolumn,superscriptaddress]{revtex4-2}
\usepackage{lipsum}
\usepackage{braket}
\usepackage{graphicx}
\usepackage{amsmath}
\usepackage{enumitem}
\usepackage[usenames,dvipsnames]{xcolor}
\usepackage[pdfencoding=auto, unicode, colorlinks=true, allcolors=blue]{hyperref}
\usepackage[capitalise]{cleveref}
\newcommand{\cX}{$\text{CX}$}
\newcommand{\cRX}{$\text{CR}_\text{X}$}

\begin{document}

\title{Hybrid cat-transmon architecture for scalable, hardware-efficient quantum error correction}

\author{Connor T.~Hann} 
\email{connohan@amazon.com}
\author{Kyungjoo Noh} 
\author{Harald Putterman} 
\author{Matthew H.~Matheny} 
\author{Joseph K.~Iverson} 
\author{Michael T.~Fang} 
\author{Christopher Chamberland}
\altaffiliation[Current affiliation: ]{Extropic}
\affiliation{AWS Center for Quantum Computing, Pasadena, CA 91125, USA}
\author{Oskar Painter} 
\affiliation{AWS Center for Quantum Computing, Pasadena, CA 91125, USA}
\affiliation{IQIM, California Institute of Technology, Pasadena, CA 91125, USA}
\affiliation{Thomas J.~Watson, Sr., Laboratory of Applied Physics,
California Institute of Technology, Pasadena, California 91125, USA}
\author{Fernando G.S.L. Brandão} 
\email{fbrandao@amazon.com}
\affiliation{AWS Center for Quantum Computing, Pasadena, CA 91125, USA}
\affiliation{IQIM, California Institute of Technology, Pasadena, CA 91125, USA}

\begin{abstract}
    Dissipative cat qubits are a promising physical platform for quantum computing, since their large noise bias can enable more hardware-efficient quantum error correction. 
    In this work we theoretically study the long-term prospects of a hybrid cat-transmon quantum computing architecture where dissipative cat qubits play the role of data qubits, and error syndromes are measured using ancillary transmon qubits. The cat qubits' noise bias enables more hardware-efficient quantum error correction, 
    and the use of transmons allows for practical, high-fidelity syndrome measurement.
    While correction of the dominant cat $Z$ errors with a repetition code has recently been demonstrated in experiment,
    here we show how the architecture can be scaled beyond a repetition code. In particular, we propose a cat-transmon entangling gate that enables the correction of residual cat $X$ errors in a thin rectangular surface code, so that logical error can be arbitrarily suppressed by increasing code distance. 
    We numerically estimate logical memory performance,
    finding significant overhead reductions in comparison to architectures without biased noise. 
    For example, with current \mbox{state-of-the-art} coherence, physical error rates of $10^{-3}$ and noise biases in the range $10^{3} - 10^{4}$ are achievable.  With this level of performance, the qubit overhead required to reach algorithmically-relevant logical error rates with the cat-transmon architecture matches that of an unbiased-noise architecture with physical error rates in the range $10^{-5} - 10^{-4}$.

\end{abstract}

\maketitle


\section{Introduction}

\label{Sec:introduction}

Bosonic quantum error-correcting codes offer a promising path towards error correction with lower qubit overhead. In contrast to \emph{qubit codes}---like the surface code~\cite{bravyi1998quantum,dennis2002topological,fowler2012surface}---that redundantly encode quantum information in the Hilbert space of multiple physical qubits, \emph{bosonic codes}~\cite{gottesman2001encoding,mirrahimi2014dynamically,michael2016new,puri2017engineering,cai2021bosonic,joshi2021quantum} encode information in the infinite-dimensional Hilbert space of a single quantum harmonic oscillator, such as an electromagnetic or mechanical resonator. The ability to already correct or suppress some errors with only one physical component makes bosonic codes naturally hardware efficient. Residual errors can then be corrected by concatenating a bosonic code with an outer qubit code. In these \emph{concatenated bosonic codes}~\cite{fukui2018high,darmawan2021practical,grimsmo2021quantum,chamberland2022building}, the reduced error of the bosonic qubits reduces the overhead to achieve a target logical error with the outer qubit code. A host of recent experiments have demonstrated the ability of various bosonic codes to correct or suppress errors, including cat codes~\cite{leghtas2015confining,ofek2016extending,touzard2018coherent,lescanne2020exponential,grimm2020stabilization,frattini2022squeezed,reglade2024quantum,hajr2024high,marquet2024autoparametric}, binomial codes~\cite{chou2018deterministic,hu2019quantum}, and GKP codes~\cite{campagne2020quantum,de2022error,sivak2023real,lachance2024autonomous}. Some experiments have even approached~\cite{hu2019quantum,campagne2020quantum,gertler2021protecting} or exceeded~\cite{sivak2023real,brock2024quantum} the so-called ``breakeven'' point, where the lifetime of the bosonic qubit exceeds the bosonic mode's coherence time. 

Among bosonic codes, cat codes are especially promising candidates for hardware-efficient error correction because they can exhibit biased noise. 
In two-component cat codes, typical physical errors, such as photon loss and dephasing, result in a probability of $X$ errors on the encoded bosonic qubit that is exponentially suppressed relative to the probability of $Z$ errors~\cite{mirrahimi2014dynamically}. 
The large ratio of the $Z$ to $X$ error probabilities, referred to as a noise bias, can then be exploited by the outer qubit code in a concatenated bosonic code.
For example, the presence of a noise bias can lead to increased error correction thresholds~\cite{aliferis2008fault,tuckett2018ultrahigh,bonilla2021xzzx,roffe2023bias} and reduced overheads~\cite{guillaud2019repetition,guillaud2021error,chamberland2022building}. 
Biased-noise cat qubits have been realized in several recent experiments, with the implementations relying on either engineered Hamiltonians~\cite{grimm2020stabilization,frattini2022squeezed,hajr2024high} or dissipation~\cite{leghtas2015confining,touzard2018coherent,lescanne2020exponential,reglade2024quantum} to stabilize the code space and achieve a noise bias. 
Notably, Refs.~\cite{lescanne2020exponential,reglade2024quantum,singlecat} demonstrate that the cat's noise bias can be increased exponentially by increasing the cat's mean photon number.

In order for the outer qubit code to take full advantage of the cats' noise bias, it is critical that this bias be presevered as gates are applied to the cat qubits. While proposals have been put forth for bias-preserving gates between cat qubits (referred to here as cat-cat gates)~\cite{puri2020bias,guillaud2019repetition}, these implementations can have onerous experimental requirements.
For example, the infidelity of Ref.~\cite{guillaud2019repetition}'s proposal for a cat-cat \cX{} gate scales as $1-F\sim \sqrt{\kappa_1/\kappa_2}$, where $\kappa_1$ is the single-photon loss rate, and $\kappa_2$ is the engineered dissipation rate of the cats~\cite{mirrahimi2014dynamically}. 
This unfortunate square-root scaling necessitates both strong engineered dissipation and high coherence for good performance.
Indeed, a recent experiment~\cite{marquet2024autoparametric} achieved an impressive $\kappa_1/\kappa_2 \approx 1/150$, but $\kappa_1/\kappa_2$ orders-of-magnitude smaller still may be needed to operate a concatenated cat code at algorithmically-relevant error rates~\cite{guillaud2021error,gouzien2023performance,chamberland2022building,le2023high}. 
It remains an open question whether such high coherence and strong engineered dissipation can be simultaneously realized in experiment. 

In this work, we propose to circumvent the unfavorable error scaling of the bias-preserving cat-cat entangling gates---while still benefiting from the cats' noise bias---by using transmon qubits, rather than cat qubits, as ancillas to measure error syndromes. 
In the resulting hybrid \emph{cat-transmon architecture}, only cat-transmon entangling gates, rather than cat-cat gates, are required to perform error correction. 
The fidelities of these cat-transmon gates scale with $\kappa_1/\chi$, where $\chi$ is the cat-transmon dispersive coupling strength. 
This scaling both enables higher fidelity gates at given $\kappa_1$, and removes the need for strong engineered dissipation.
Moreover, because the gates' implementations rely on the natural qubit-oscillator dispersive interaction, there is no need for complicated Hamiltonian engineering, and fast gates are enabled by easily-engineered MHz-scale dispersive couplings. 
These practical benefits allow for fast, high-fidelity cat-transmon gates to be readily implemented, enabling near-term experimental demonstrations of concatenated cat codes, including the recent realization of a repetition cat code in Ref.~\cite{repcat}.

The price we pay for these practical benefits is an upper limit on the achievable noise bias. This limit is due to the propagation of transmon ancilla errors into cat data qubit $X$ errors via the cat-transmon gates. 
Steps are taken to mitigate this error propagation~\cite{rosenblum2018fault,reinhold2020error,ma2020path}, such that no single transmon loss or dephasing error can induce a cat $X$ error on its own. We project that significant biases, in the range of $10^3-10^4$, are achievable with current state-of-the-art parameters as a result. 
Even so, a way to correct the residual cat $X$ errors is still necessary for the cat-transmon architecture to be fully scalable, i.e.~to achieve arbitrarily low logical error. We propose a cat-transmon entangling gate, the \cRX{} gate, that enables this correction while enjoying all of the practical benefits mentioned above. 

With this blueprint for a scalable architecture in hand, we proceed to quantify the hardware efficiency of the cat-transmon architecture relative to a generic unbiased-noise architecture. We find that significant reductions in logical memory overhead are possible despite the limited bias. 
For example, physical error rates of $10^{-3}$ are accessible with current state-of-the-art coherence, $\kappa_1/\chi \sim 10^{-5}-10^{-4}$. With this level of error, we project that a thin rectangular surface code with logical memory error $<10^{-10}$ could be built with only $200$ qubits (100 cats, 100 transmons), thanks to the sizable noise bias ($10^3-10^4$) of the cat-transmon architecture. In contrast, with the same physical error but without noise bias, nearly 1000 qubits would be required to construct a comparably performant square surface code. Alternatively, without bias, physical error rates would need to be reduced to $<10^{-4}$ to match the qubit overhead of the cat-transmon architecture.
Of course, while the larger biases that cat-cat gates offer could yield even further improvement, we emphasize that the primary appeal of the cat-transmon architecture is its practicality; significant noise bias and hardware efficiency are achievable using only well-established experimental platforms and capabilities. 

The rest of this manuscript is organized as follows. In \cref{Sec:architecture} we provide a high-level overview of the cat-transmon architecture, focusing on how different cat-transmon gates enable the correction of both the dominant cat $Z$ errors and the suppressed cat $X$ errors. Next, in \cref{Sec:cRX} we present a detailed description of the bias-preserving \cRX{} gate that enables correction of the suppressed cat $X$ errors. We numerically benchmark the performance of these gates in \cref{Sec:achievable}, and then we use the results to make projections for the cat-transmon architecture's performance as a logical memory in \cref{Sec:logical}. Finally, in \cref{sec:discussion} we conclude by discussing several directions for future research to further improve the performance of the architecture.


\section{Hybrid cat-transmon architecture}
\label{Sec:architecture}

\begin{figure*}
    \centering
    \includegraphics[width=\textwidth]{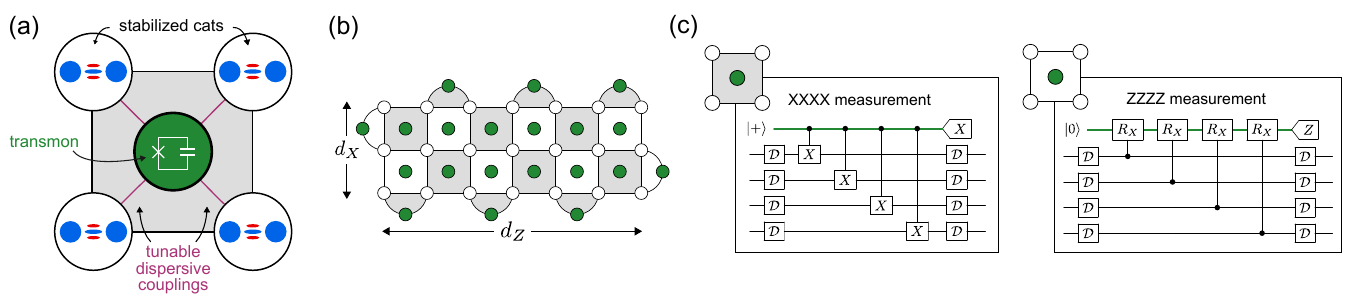}
    \caption{Hybrid dissipative cat-transmon QEC architecture. \textbf{(a)} Cat-transmon unit cell. Dissipatively stabilized cats act as data qubits, and  transmons are employed to extract error syndromes. The cats and transmons are coupled via tunable dispersive couplings. \textbf{(b)} Surface codes built from the cat-transmon unit cell. By constructing rectangular codes that offer more protection against the cats' dominant $Z$ errors than their suppressed $X$ errors, the cats' biased noise can allow for more hardware-efficient error correction. \textbf{(c)} Syndrome measurement circuits. To measure $X$-type syndromes, transmon-controlled \cX{} gates are implemented using the native dispersive couplings. To measure $Z$-type syndromes, the architecture employs a proposed cat-controlled entangling gate, termed \cRX{}. To prevent leakage out of the cat-qubit code space, engineered dissipation is applied to the cats during the transmon ancilla readout and reset steps (indicated by boxes labelled “$\mathcal{D}$”).}
    \label{fig:architecture}
\end{figure*}

In this section, we provide an overview of the cat-transmon architecture (\cref{fig:architecture}). We first discuss the architecture's basic building blocks: the dissipatively-stabilized cat data qubits and the dispersively-coupled transmon ancilla qubits. Then, we describe how cat-transmon entangling gates enable error correction in an outer surface code.

\subsection{Physical components}

The data qubits in our architecture are encoded in two-component cat codes, which store information in superpositions of coherent states $\ket{\pm \alpha}$. The logical $X$ basis states of the cat code, $\ket{\pm}_C$, are defined as the even- and odd-parity superpositions of these coherent states
\begin{equation}
    \ket{\pm}_C = \mathcal{N}_\pm(\ket{\alpha} \pm \ket{-\alpha})
\end{equation}
where $\mathcal{N}_\pm = 1/\sqrt{2(1\pm e^{-2 |\alpha|^2})}$ are normalization constants. The logical $Z$ basis states of the cat code, $\ket{0/1}_C$, are then given by the superpositions 
\begin{align}
\ket{0/1}_C &= \frac{1}{\sqrt{2}}(\ket{+}_C \pm \ket{-}_C) \nonumber
\\
 &= \ket{\pm \alpha} + O(e^{-2|\alpha|^2}) \ket{\mp \alpha},    
\end{align}
with $\ket{0/1}_C$ approaching the coherent states $\ket{\pm \alpha}$ in the limit of large $|\alpha|^2$. We refer to the bosonic modes that host cat qubits in our architecture as ``storage modes.''

When the state of a storage mode is confined to the cat code space, the cat qubit exhibits biased noise. 
In this work, we focus on the paradigm of dissipative stabilization: confinement to the code space is engineered by a driven-dissipative process where the storage mode exchanges pairs of photons with its environment, which can be engineered, e.g., by nonlinearly coupling the mode to a separate lossy mode~\cite{mirrahimi2014dynamically,leghtas2015confining,touzard2018coherent,lescanne2020exponential}. The dynamics of the dissipatively stabilized system can be described by the Lindblad master equation
\begin{equation}
    \dot \rho(t) = \kappa_2\mathcal{D}[a^2-\alpha^2]\rho(t)
    \label{eq:engineered_dissipation}
\end{equation}
where $a$ is the storage mode's annihilation operator, $\kappa_2$ is the engineered dissipation rate, and $D[L]\rho(t) = L\rho L^\dagger -(L^\dagger L \rho + \rho L^\dagger L)/2$. The steady state manifold of these dynamics is the cat code space. Once confined to the code space, typical error mechanisms like single-photon loss ($L = \sqrt{\kappa_1} a$) and white-noise dephasing ($L=\sqrt{\kappa_\phi} a^\dagger a$) result in a noise bias that grows exponentially with $|\alpha|^2$. For example, single-photon loss will induce cat $X$ errors at a rate
$\propto e^{-c |\alpha|^2}$, where the exponential prefactor $c$ depends on the parameter regime \cite{dubovitskii2024bit}, and $Z$ errors at a rate $\propto |\alpha|^2$. The exponential suppression of the $X$ error rate with $|\alpha|^2$ relative to the linear increase in the $Z$ error rate results in a tunable bias that can be deliberately made large, as demonstrated in recent experiments~\cite{lescanne2020exponential,berdou2022one,marquet2024autoparametric,reglade2024quantum}.  We note that the Kerr cat~\cite{puri2017engineering} also constitutes a promising paradigm for cat stabilization, as demonstrated in recent experiments~\cite{grimm2020stabilization,frattini2022squeezed,hajr2024high}. However, here we restrict our attention to dissipative cats, since large Kerr nonlinearities are not easily compatible with cat-transmon gate implementations discussed below. 

The ancillary qubits in our architecture are standard transmon qubits and are coupled dispersively to the storage modes.
The dispersive interaction is described by an effective Hamiltonian of the form~\cite{blais2021circuit}
\begin{equation}
    H = \chi_e \ket{e}\bra{e} a^\dagger a + \chi_f \ket{f}\bra{f} a^\dagger a
    \label{eq:dispersive_coupling}
\end{equation}
where $\ket{e}$ and $\ket{f}$ denote the transmon's first and second excited states, and $\chi_{e,f}$ denote the corresponding dispersive frequency shifts of resonator. We will exploit the fact that a qubit can be encoded in the subspace spanned by the transmon's ground and first excited states, $\ket{g}$ and $\ket{e}$, or alternatively in its ground and second excited states, $\ket{g}$ and $\ket{f}$. 

Our implementations of cat-transmon gates place certain requirements on the device parameters. First, we require that both the engineered dissipation strength $\kappa_2$ and the dispersive couplings $\chi$ can be quickly toggled on and off, with $\chi^{(\text{on})} \gg \kappa_{2}^\text{(off)}$ and $\kappa_{2}^\text{(on)} \gg \chi^\text{(off)}$. Such tunability has been demonstrated in recent experiments, where, for example, $\kappa_2$ can be controlled via external flux drive~\cite{lescanne2020exponential}, and $\chi$ can be controlled via a tunable coupler~\cite{singlecat}. Second, when the transmon is operated in the $\ket{g},\ket{f}$ manifold, we require $\chi_e \approx \chi_f$. This approximate ``chi matching'' is required to achieve a large noise bias, and can be realized either actively via Hamiltonian engineering~\cite{rosenblum2018fault,reinhold2020error,ma2020path}, or passively in design via a careful arrangement of underlying device parameters~\cite{repcat}. We emphasize that the chi matching here needs only to be approximate, as the cat's dissipative stabilization can correct the effects of a small chi mismatch, in contrast to earlier experiments~\cite{rosenblum2018fault,reinhold2020error} that lacked dissipative stabilization and so required essentially perfect matching.  Beyond these requirements, we do not otherwise specify implementation details of how the dissipative stabilization or dispersive interactions are engineered, so that the results below are not restricted to a particular implementation. That said, we note that Ref.~\cite{repcat} constitutes a concrete example of an architecture that simultaneously satisfies all of these requirements.

\subsection{Gates and error correction}

The basic unit cell of our architecture, the cat-transmon unit cell, is shown in \cref{fig:architecture}(a). Surface codes can be obtained by tiling the cat-transmon unit cell, as shown in \cref{fig:architecture}(b).
Importantly, we employ rectangular codes to capitalize on the cats' biased noise.
Rectangular codes are parameterized by two different code distances, $d_X$ and $d_Z$, that separately characterize the degree of protection against $X$ and $Z$ errors, respectively. Because the cat qubits' $X$ errors are already strongly suppressed, it suffices to use relatively small $d_X$, reducing the qubit overhead relative to a square code. 
We note that there are other ways to leverage noise bias to improve hardware efficiency (e.g.,~by employing Clifford-deformed surface codes like the XZZX~\cite{bonilla2021xzzx} or XY~\cite{tuckett2018ultrahigh} codes), but we focus on the standard (i.e.~rotated, non-deformed) rectangular surface code because 
we find that it offers a larger hardware-efficiency improvement once errors are far below threshold (see \cref{App:code_comparison} for analysis of the XZZX code and comparison). 

Circuits for measuring the surface code stabilizers are shown in \cref{fig:architecture}(c). The two types of stabilizer, $X$-type and $Z$-type, enable the correction of $Z$ and $X$ errors, respectively. 
Correcting both types of errors is crucial for full scalability, given the finite noise bias.
Two distinct gates are required to measure the syndromes, both because we use different implementations for the data and ancilla qubits, and also because the use of operations that convert $Z$ errors to $X$ errors, like Hadamard gates, must be avoided on the cat qubits to maintain their noise bias.   The $X$-type syndrome measurement requires a transmon-controlled $X$ operation on the cat, while the $Z$-type measurement requires a cat-controlled $X$ operation on the transmon.

To correct the cats' dominant $Z$ errors, a transmon-controlled \cX{} gate is implemented following the approach of Ref~\cite{rosenblum2018fault}. Entanglement between the cat and transmon is generated simply by letting the system evolve freely under the dispersive coupling, with the cat's engineered dissipation turned off. For a transmon encoded in the $\ket{g},\ket{e}$ (resp. $\ket{g},\ket{f}$) manifold, evolving for a time $\pi/\chi_e$ $(\pi/\chi_f)$ results in the storage mode acquiring a $\pi$ phase shift conditioned on the transmon being in the excited state. This phase shift effectively swaps $\ket{+\alpha} \leftrightarrow \ket{-\alpha}$, equivalent to a transmon-controlled \cX{} gate.  

If the transmon is operated in the $\ket{g},\ket{e}$ manifold, the cat's noise bias is substantially compromised by this transmon-controlled \cX{} gate, but if the transmon is operated instead in the $\ket{g},\ket{f}$ manifold~\cite{rosenblum2018fault,reinhold2020error,ma2020path}, a large noise bias can be retained.
For $\ket{g},\ket{e}$, the bias is compromised because single transmon errors can readily propagate to cat $X$ errors. For example, if the transmon suffers a $T_1$ decay error, $\ket{e} \to \ket{g}$, immediately before the gate, the storage mode does not acquire the desired phase shift, amounting to a cat $X$ error. This problem can be mitigated by instead operating the transmon in the $\ket{g}, \ket{f}$ manifold while employing chi matching $\chi_e \approx \chi_f$. Now, the resonator acquires the desired phase shift even if the transmon decays from $\ket{f} \to \ket{e}$ during the gate. This error transparency~\cite{ma2020path,ma2022algebraic,xu2024fault} means that single transmon decay errors do not propagate to cat $X$ errors. However, less probable double-decay ($\ket{f} \to \ket{e} \to \ket{g}$) or heating ($\ket{g}\to\ket{e}$) errors can still propagate to cat $X$ errors, and these errors place an upper limit on the achievable bias. We thus refer to this transmon-controlled \cX{} as only \emph{moderately} noise biased, and we reserve the term \emph{exponentially} noise biased for implementations where the cat's bias grows arbitrarily and exponentially with $|\alpha|^2$. In \cref{Sec:achievable}, we numerically demonstrate that significant bias is nevertheless achievable with this \cX{} gate. 

To correct the cats' suppressed $X$ errors, we propose an implementation of a cat-controlled transmon $X$ rotation, referred to as the \cRX{}, that is exponentially noise biased. The \cRX{} gate is implemented via a composite pulse sequence, involving monochromatic drives applied to the transmon and storage mode, where again entanglement is generated by the dispersive coupling and engineered dissipation is turned off during the gate. Notably, the \cRX{} gate can be exponentially noise biased, despite only requiring operating the transmon in the $\ket{g},\ket{e}$ manifold.  We discuss the implementation in detail in \cref{Sec:cRX}, and we show that the gate is exponentially noise biased in \cref{Sec:achievable}.

Having summarized the \cX{} and \cRX{} implementations, we highlight the practical appeal of these gates. Both gates rely on native dispersive coupling to generate entanglement, so complicated Hamiltonian engineering is not required. This simplicity has already enabled multiple experimental demonstrations of the \cX{} gate~\cite{rosenblum2018fault,reinhold2020error,repcat}, along with a demonstration of a rudimentary \cRX{} in Ref.~\cite{repcat} (where it was used to enable $Z$-basis cat readout, see \cref{app:crx_applications}). Moreover, both gates also avoid the use of engineered dissipation, so that the dissipation strength does not contribute to gate fidelity. Instead, as observed in Ref.~\cite{gautier2022combined}, it suffices to apply the engineered dissipation only intermittently, while the cat qubits idle [see \cref{fig:architecture}(c)]. 
As demonstrated experimentally in Ref.~\cite{singlecat,repcat}, the required dissipation strength to preserve $X$ error suppression with this ``pulsed'' dissipation is quite modest (the requirement on dissipation strength is quantified in \cref{app:pulsed_stabilization}). 
Both the simplicity of implementation and modest requirement on engineered dissipation strength for these cat-transmon gates are points of contrast with known implementations of cat-cat gates, and as we show in \cref{Sec:achievable} the former offer high fidelity and significant bias with current state-of-the-art parameters. 


\section{\texorpdfstring{$\text{CR}_\text{X}$}{CRX} gate: cat-controlled transmon rotation}
\label{Sec:cRX}

In order to measure $Z$-type syndromes and hence correct cat $X$ errors, we require a cat-controlled entangling gate. For this purpose, we propose a gate implementation, dubbed the \cRX{} gate, where the transmon and cat are entangled via an alternating sequence of storage and transmon drives. The gate is exponentially noise biased, and incorporates both dynamical decoupling and pulse shaping to minimize error. The \cRX{} gate builds on ideas from the gate implementations in Refs.~\cite{leghtas2013hardware, eickbusch2022fast, diringer2023conditional, xu2024fault} to realize an exponentially noise-biased cat-controlled gate.

In the following subsections, we first describe the gate's implementation, then provide further detail on two aspects---storage echoing and selective pulse shaping---that are crucial to achieving high fidelity. Numerical simulations of the gate are presented in \cref{Sec:achievable}. Additionally, while here we focus on the \cRX{} gate's use to correct cat $X$ errors, in  \cref{app:crx_applications} we discuss other applications of the gate that may be of independent interest. These include realizations of high-fidelity cat $Z$ and CZ gates (i.e.~with infidelity scaling $\sim\kappa_1/\chi$ rather than $\sim\sqrt{\kappa_1/\kappa_2}$), as well as single-shot $Z$-basis cat readout with readout error exponentially suppressed in $|\alpha|^2$

\subsection{Pulse sequence}

\begin{figure}
    \centering
    \includegraphics[width=0.9\columnwidth]{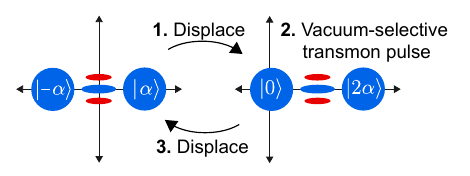}
    \caption{Simplified schematic of the \cRX{} gate. To implement the cat-controlled transmon rotation, the cat is first displaced by $+\alpha$, such that $\ket{-\alpha}\to\ket{0}$. A number-selective drive is then applied to the transmon, with the drive frequency resonant with the transmon frequency when the storage is in vacuum. The storage mode is subsequently displaced back, and the net effect is a rotation of the transmon that occurs only if the cat was initially in $\ket{-\alpha}$. Note that this simplified diagram neglects transmon-dependent rotation of the storage mode during the selective pulse due to the dispersive coupling.  }
    \label{fig:cRX_idea}
\end{figure}

The main idea behind the gate is sketched in \cref{fig:cRX_idea}. First, a drive on the storage mode is applied to displace the mode by $+\alpha$, such that the coherent states $\ket{\alpha}$ and $\ket{-\alpha}$ are mapped $\ket{2\alpha}$ and $\ket{0}$ respectively. Next, a \emph{number-selective} transmon pulse is applied, such that the transmon is flipped from $\ket{g} \leftrightarrow \ket{e}$ only if the storage is in $\ket{0}$, and the transmon is ideally unaffected otherwise. This selectivity is enabled by the cat-transmon dispersive coupling (Eq.~\ref{eq:dispersive_coupling}). This coupling shifts the transmon frequency conditioned on the number of photons in the storage, so a number-selective pulse is achieved simply by applying a frequency-selective pulse to the transmon. Finally, the storage mode is displaced back by $-\alpha$, returning the system to the cat code space. 

The net result of this procedure is to flip the state of the transmon only if the cat qubit was initially in the state $\ket{-\alpha}$. More precisely, the gate implements the unitary
\begin{equation}
    \text{\cRX{}} = \ket{\alpha}\bra{\alpha} \otimes I + \ket{-\alpha}\bra{-\alpha} \otimes R_X(\pi),
    \label{eq:cRX}
\end{equation}
where $R_X(\pi) = e^{-i \pi X /2 }$, with $X = \ket{g}\bra{e} + H.c.$. We refer to this unitary as the \cRX{} gate. The \cRX{} is locally equivalent to a cat-controlled \cX{} gate, the difference being that an additional phase gate, $S^\dagger = \mathrm{diag}(1,-i)$, is applied to the cat by the \cRX{}. This local phase must be accounted for but does not inhibit our ability to use the \cRX{} for syndrome measurements (see \cref{App:cRX_phase}).

Why is the \cRX{} gate exponentially noise biased? The intuition is that a cat $X$ error requires a large unintentional rotation of the storage mode. If the \cRX{} gate's constituent pulses can be applied quickly enough, such that there is not sufficient time for the cat-transmon dispersive coupling to generate such a large rotation, then the probability of a cat $X$ error will be minimal. Indeed, numerical simulations of the gate (discussed further below) indicate that applying the pulses in a time $T_\mathrm{cR_X} \leq \pi/\chi$ is sufficient to achieve a low cat $X$ error probability. We emphasize that, in contrast to the \cX{} gate, the \cRX{} gate does not require that we use higher transmon levels to achieve a large cat noise bias. 

The above description of the \cRX{} is a simplification. 
In practice, inserting additional pulses is both necessary to keep the cat phase well defined and beneficial to protect against dephasing as discussed below.
The gate sequence we actually propose is as follows:
\begin{enumerate}
    \itemsep0em 
    \item Storage displacement: $D(+\alpha)$
    \item Selective transmon pulse: $R_X(\pi/4)$
    \item Unselective transmon pulse: $R_X(\pi)$
    \item Selective transmon pulse: $R_X(\pi/4)$
    \item Storage displacement: $D(-2\alpha e^{i\phi})$
    \item Selective transmon pulse: $R_X(-\pi/4)$
    \item Unselective transmon pulse: $R_X(-\pi)$
    \item Selective transmon pulse: $R_X(-\pi/4)$
    \item Storage displacement: $D(\alpha e^{2i\phi})$
\end{enumerate}
Here, $D(\alpha)$ denotes a storage displacement with amplitude $\alpha$, and $\phi = \chi T_\mathrm{sel.}$ where $T_\mathrm{sel.}$ denotes the duration of a selective $R_X(\pm\pi/4)$ pulse. 
We emphasize that though the sequence comprises nine different pulses, all of the constituent pulses are routine and easily implementable. 
In \cref{App:cRX_details}, we show that this sequence is equivalent to the \cRX{} up to an $R_X(\pi/2)$ rotation on the transmon, and we also show that, if desired, the gate can be reduced to five pulses at the price of increased sensitivity to dephasing.
We note also that by substituting the selective $R_X(\pi/4)$ rotations with $R_X(\theta/4)$ rotations, the above pulse sequence also suffices to implement a general cat-controlled $R_X(\theta)$ rotation of the transmon.

\begin{figure*}[htbp]
    \centering
    \includegraphics[width=\textwidth]{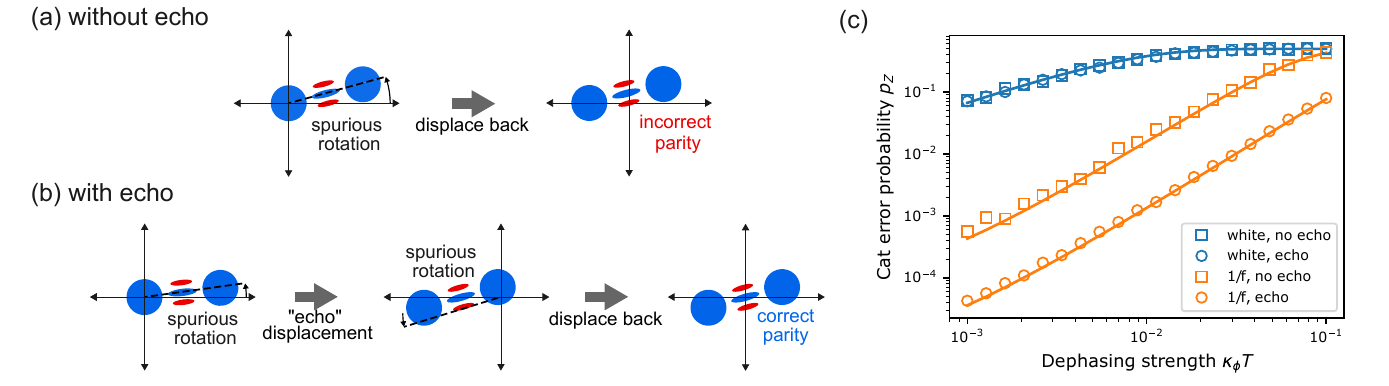}
    \caption{Storage ``echo'' to mitigate dephasing-induced phase flips during the \cRX{} gate. \textbf{(a)} Without the echo displacement, spurious rotations in a displaced cat can result in parity flips (cat $Z$ errors) once displaced back, graphically indicated by the positioning of the red interference fringe at the origin in the right diagram. \textbf{(b)} To mitigate these $Z$ errors, an “echo displacement” is applied in between successive selective pulses, such that the cat spends equal time on the left and right halves of phase space. \textbf{(c)} The final probability of a cat Z error is plotted as a function of the storage dephasing strength for both white and $1/f$ dephasing noise. The inclusion of the storage echo significantly reduces the error probability for the low-frequency dominated $1/f$ noise, but not for white noise. Markers indicate trajectory simulation results, and solid lines are analytical predictions. See \cref{App:storage_echo} for derivations and numerical details.  }
    \label{fig:storage_echo}
\end{figure*}

\subsection{Mitigating dephasing with transmon and storage echoes}

The pulse sequence incorporates echoes on both the transmon and storage in order to protect against low-frequency dephasing noise. The transmon echoes, implemented via the two unselective transmon $\pi$ pulses (steps 3 and 7), are a standard technique to suppress the impacts of low-frequency dephasing noise~\cite{bylander2011noise}. Such echoes are similarly employed in the echoed conditional displacement gate~\cite{eickbusch2022fast}, for example.

Importantly, though, in the case of the \cRX{} gate these transmon $\pi$ pulses are necessary to ensure that the phase of the cat is well defined, as in Ref.~\cite{leghtas2013hardware}. During the selective pulses, the dispersive coupling causes the storage mode to accumulate phase conditioned on the transmon state. The $\pi$ pulse at step 3 (7) ensures that, absent transmon decay or heating, the storage mode accumulates a fixed phase $\phi = \chi T_\mathrm{sel}$ over steps 2-4 (6-8) that is independent of the initial transmon state.  Without echoes, the storage phase accumulated by the end of the gate would be dependent on the transmon state, making the final displacement back to the code space (step 3 of \cref{fig:cRX_idea}) impossible. With echoes, the total phase accumulated by the storage mode is fixed at $2\phi$, and this phase accumulation can be tracked classically in software.

The ``storage echo,'' implemented via the intermediate displacement (step 5), is a nonstandard technique to suppress the impacts of low-frequency storage dephasing, and we illustrate the idea conceptually in \cref{fig:storage_echo}(a,b). Photon-number-preserving perturbations, like dephasing or self-Kerr, do not normally induce cat $Z$ errors, since they commute with the cat $X$ operator (the photon number parity operator). However, these same perturbations can induce $Z$ errors in a displaced cat. Indeed, in the displaced frame the dephasing operator becomes $O_\text{dephase} = a^\dagger a \to(a^\dagger + \alpha^*)(a + \alpha)$, which contains terms that do not preserve photon number parity. To mitigate these errors, the storage echo displaces the cat to the opposite side of phase space midway through the gate, so that on average the parity-non-preserving components of perturbations like low-frequency dephasing or self-Kerr cancel. Explicitly, one can see this cancellation by moving to the interaction picture with respect to the displacements, then computing the time-averaged dephasing operator $\overline{O}_\text{dephase}$ (i.e.~a leading-order Magnus expansion~\cite{magnus1954exponential}),
\begin{align}
    \overline{O}_\text{dephase} &\approx 
    \frac{1}{2}\left[(a^\dagger + \alpha^*)(a + \alpha) + (a^\dagger - \alpha^*)(a - \alpha) \right] \nonumber \\
    &= a^\dagger a +\text{const.}
\end{align}

We demonstrate the efficacy of this storage echoing technique in \cref{fig:storage_echo}(c). Following Ref.~\cite{cywinski2008enhance}, in \cref{App:cRX_details}, we show that the cat $Z$ error probability as a function of time is given by $p_Z(t) = 1- \frac{1}{2}(1+e^{-\Gamma_\phi(t)})$, where
\begin{equation}
\label{eq:dephasing_rate}
    \Gamma_\phi(t) = 16 |\alpha|^4 \int_0^\infty \frac{d\omega}{2\pi} S(\omega) \frac{F(\omega t)}{\omega^2}.
\end{equation}
Here $S(\omega)$ frequency-noise power spectral density, and $F(\omega t)$ is a filter function that takes different forms depending on whether the echo is applied. Without the echo $F(\omega t) = 2 \sin^2(\omega t/2)$, and with the echo $F(\omega t) = 8 \sin^4(\omega t/2)$. In \cref{fig:storage_echo}(c) we plot these analytical expressions for the dephasing rate alongside the results of numerical simulations incorporating dephasing noise with either white-noise ($S(\omega)=$ constant) or $1/f$ ($S(\omega)\propto1/\omega$) spectra~\cite{didier2019ac} (see \cref{App:storage_echo}). As expected, we find that the storage echo significantly reduces in the probability of a cat $Z$ error due to the low-frequency-dominated  $1/f$ dephasing noise but does not offer any protection against white-noise dephasing.

\subsection{Mitigating coherent error with pulse shaping}
\label{sec:pulse_shaping}

\begin{figure}
    \centering
    \includegraphics[width=0.9\columnwidth]{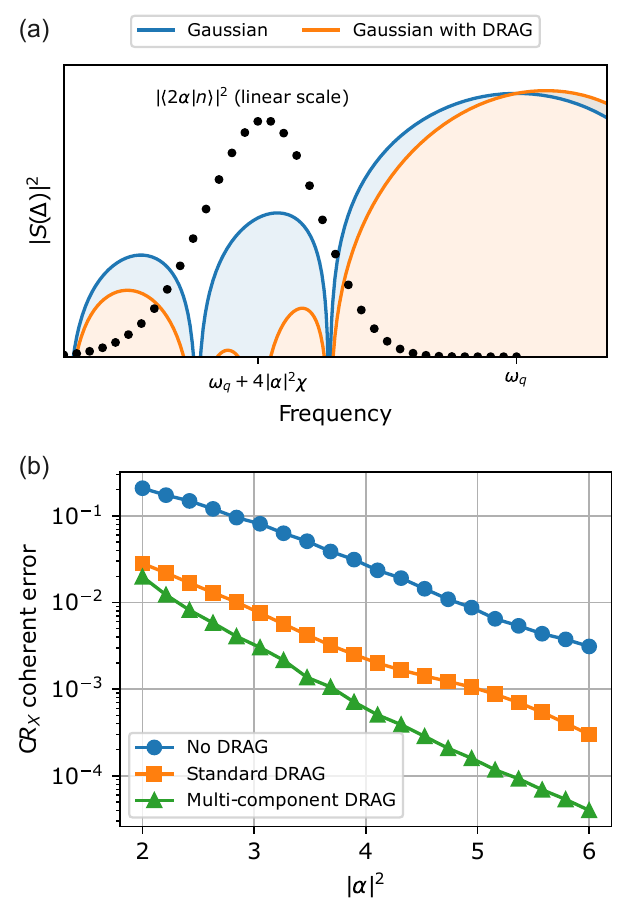}
    \caption{Mitigating \cRX{} coherent error with pulse shaping. \textbf{(a)} Frequency spectra of standard and shaped pulses. The shaped DRAG pulse is designed to minimize spectral content at the frequency of the transmon when $|2\alpha|^2$ photons are in the storage, i.e.~$\omega_q+2|\alpha|^2 \chi$, where $\omega_q$ is the unshifted transmon frequency. To guide the eye, the photon number distribution for the coherent state $|2\alpha\rangle$ is plotted (on a linear scale, not shown), with $|\braket{2\alpha|n}|^2$ positioned at $\omega_q+n\chi$ on the horizontal axis. At the frequencies where $|2\alpha\rangle$ has significant support, the spectral content of the DRAG pulse is significantly suppressed relative to the case without DRAG. \textbf{(b)} Coherent error reduction for via DRAG. The coherent error of different selective pulse ansatzes is plotted as a function of $|\alpha|^2$, where at each point the free parameters of a pulse ansatz are optimized to minimize coherent error. The coherent error is computed as the average gate infidelity of the Pauli error channel extracted from simulations of the \cRX{} gate without decoherence (see \cref{app:extracting_error_channels}).  }
    \label{fig:DRAG}
\end{figure}

Coherent errors arise in the \cRX{} gate due to the finite duration of the selective transmon pulses. As an example, consider step 2 of the gate, where an ideal selective pulse should enact a $R_X(\pi/4)$ rotation of the transmon if the storage mode is in $\ket{0}$, and act trivially on the transmon if the storage mode is in $\ket{2\alpha}$. In the former case, the transmon frequency is not dispersively shifted and takes its bare value $\omega_0$. In the latter case, there is an indeterminate dispersive shift; the transmon frequency is $\omega_0 + n \chi$, where the probability of having $n$ photons in the storage is $p(n) = |\langle  2\alpha |n\rangle|^2$. 
Coherent errors necessarily arise in finite-duration selective pulses because the desired process at $\omega_0$ cannot be fully spectrally resolved from undesired processes detuned by $n \chi$, which is especially challenging at small cat mean photon number.  The probability of an unwanted transition process detuned by $\Delta$ is related to $p(n = \Delta/\chi)$, together with the spectral content of the selective pulse shape, $\Omega(t)$, at the corresponding detuning, $|S_\Omega(\Delta)|^2$, where $S_\Omega(\Delta) = \int dt \Omega(t) e^{i \Delta t}$.

Conveniently, these coherent errors can be easily suppressed through pulse shaping. \cref{fig:DRAG}(a) conceptually illustrates how a simple DRAG pulse~\cite{motzoi2009simple, krantz2019quantum} mitigates these errors. The DRAG pulse augments the original pulse $\Omega_0(t)$ by adding an out-of-phase component proportional to its derivative, $\Omega(t) = \Omega_0(t) - \beta i \dot \Omega_0(t)$, where the constant $\beta$ is specifically chosen to suppress spectral content at $\Delta = |2\alpha|^2 \chi$, i.e.~at the photon number where $p(n)$ is maximal. \cref{fig:DRAG}(b) numerically illustrates the efficacy of this simple technique. Replacing a simple Gaussian $\Omega_0(t)$ with this ``standard'' DRAG pulse results in nearly an order-of-magnitude reduction in the coherent error of a \cRX{} gate. 

The standard DRAG protocol is designed to suppress an unwanted transition at only a single frequency, but the \cRX{} coherent error can be even further reduced by generalized DRAG pulses that suppress multiple unwanted transitions simultaneously~\cite{motzoi2013improving,xu2022engineering}. For example, the pulse
\begin{equation}
    \Omega(t) = \Omega_0(t) -i \left[\frac{1}{\Delta_1}+\frac{1}{\Delta_2}\right]\dot \Omega_0(t) - \frac{1}{\Delta_1\Delta_2}\ddot{\Omega}_0(t)
\end{equation}
approximately suppresses transitions at two frequencies, $\Delta_1$ and $\Delta_2$; see \cref{App:cRX_details} for derivation. By choosing to distribute the $\Delta_1$ and $\Delta_2$ of  over the range of frequencies where $p(n)$ appreciable, it is possible to achieve significantly lower coherent error with this ``multi-component DRAG'' than with standard DRAG, as demonstrated in \cref{fig:DRAG}(b). Moreover, while in general DRAG's addition of out-of-phase components can degrade the fidelity of the desired transition, in \cref{App:cRX_details} we show that this degradation can be entirely avoided in the context of the \cRX{} gate. In particular, adding a time-dependent pulse detuning,
\begin{equation}
\delta(t) =- \Re\Omega(t)\tan\left[\int_0^tdt' \Im\Omega(dt')\right] 
\end{equation}
and adjusting the amplitude of $\Omega(t)$ to satisfy
\begin{equation}
\theta=\int_0^Tdt\,\Re\Omega(t)\sec\left[\int_0^{t} dt' \Im \Omega(t)\right]
\end{equation}
exactly implements an $R_X(\theta)$ rotation at the desired transition, regardless of any DRAG correction.  Pulse shaping thus enables negligible \cRX{} coherent error even at modest cat mean photon numbers. For example, in \cref{fig:DRAG}(b) the multi-component DRAG pulses achieve coherent error $<10^{-3}$ already for $|\alpha|^2\gtrsim 4$. Even so, the residual \cRX{} coherent error disincentivizes operation at small $|\alpha|^2$. See \cref{App:cRX_details} for derivations and details on the numerics.


\section{Error and bias of cat-transmon gates}
\label{Sec:achievable}

In this section, we numerically benchmark the performance of the two entangling gates, \cX{} and \cRX{}, that are required for error correction in the cat-transmon architecture. Despite the limitations imposed by transmon decoherence, we find that the cat-transmon gates can be implemented with high fidelity and bias for currently-achievable coupling strengths and coherence times. We begin by first describing the noise model used in the simulations then present projections for the fidelity and bias of the two gates.

\subsection{Noise model and assumptions}

The noise model used in simulations of the cat-transmon gates is summarized in \cref{tab:noise_model}. The model is parametrized by a single dimensionless parameter, $q$, that serves as a measure of decoherence in both the storage mode and transmon. In our gate simulations, the only physically relevant quantities are ratios of decoherence rates to $\chi$, since both the \cX{} and \cRX{} gate times scale inversely with $\chi$, so these ratios are all specified in terms of $q$. This single-parameter model is both convenient to analyze and embodies the fact that, in practice, the coherence of resonators and transmons in a device often vary together as a function of the quality of materials and fabrication. 

For context, values of $q$ in the range of $10^{-4}-10^{-5}$ are realistically achievable with state-of-the-art devices in the near term. 
For example, the device of Ref.~\cite{chou2023demonstrating} simultaneously exhibited near-millisecond storage $T_1 =\kappa_1^{-1}$ and MHz-scale $\chi$, achieving $\kappa_1/\chi = q\approx10^{-4}$. With few-millisecond storage $T_1$, as has already been demonstrated in similar devices~\cite{reagor2016quantum}, $q$ closer to $10^{-5}$ is achievable.
Realistic values of $q$ for near-term 2D devices are higher by roughly an order of magnitude, $q \sim 10^{-3}-10^{-4}$, with $q\approx 10^{-3}$ representative of the device in Ref.~\cite{repcat} as an example.
In the long term, transmon coherence could limit the ability of the cat-transmon architecture to capitalize on ultrahigh intrinsic storage $T_1$ ($>10$ ms~\cite{milul2023superconducting}); expectations for achievable $q$ in the long term for both 2D and 3D platforms are discussed in \cref{sec:discussion}.

We note some of the physical considerations that motivate the specific values in \cref{tab:noise_model}. The model assumes that the storage mode $T_1 = \kappa_1^{-1}$ is 3 times larger than the transmon $T_1 = \gamma_\downarrow^{-1}$, which is roughly representative of recent 3D devices~\cite{sivak2023real, chou2023demonstrating} as well as the 2D device of Ref.~\cite{repcat}. This separation between storage and transmon lifetimes may be less representative of 2D devices generally, however, so in  \cref{app:alternative_noise_models} we consider an alternative noise model where storage and transmon lifetimes are comparable.   For the transmon, the model of \cref{tab:noise_model} assumes an effective thermal population of $\gamma_\uparrow/\gamma_\downarrow =0.5\%$, and $T_1 = T_2$, where $T_1 = \gamma_\downarrow^{-1}$ and $T_2 = (\gamma_\downarrow/2 + \gamma_\phi)^{-1}$, both of which are reasonable assumptions for typical transmon devices. For the storage mode, while dephasing noise spectra of superconducting resonators have been observed to exhibit a $1/f$ frequency dependence~\cite{murch20121}, we simulate white- rather than $1/f$-dephasing noise for the sake of numerical efficiency. The choice of a small white-noise pure dephasing rate $\kappa_\phi \ll \kappa_1$ is then motivated by two factors: (1) it is representative of 3D cavities, which exhibit little intrinsic dephasing~\cite{eickbusch2022fast,milul2023superconducting}, and (2) as \cref{fig:storage_echo} shows, the choice of small white noise dephasing rate ($\kappa_\phi \ll \kappa_1$) is representative of a much larger $1/f$ dephasing rate ($\kappa_\phi \sim \kappa_1$) in terms of the impact on gate performance. 

Finally, we emphasize that our simulations assume the cat-transmon interaction is described by the simple, effective dispersive coupling Hamiltonian in \cref{eq:dispersive_coupling}. 
In practice, to produce quantitative agreement with a specific experimental realization of a cat-transmon system, it would likely be necessary to employ a more detailed model that incorporates, e.g., details of how the tunable coupling and dissipative stabilization are realized.
However, we deliberately adopt this simple effective model so that we can assess the potential of the proposed architecture generally, without restricting our focus to a particular hardware realization. See \cref{App:sim_assumptions} for further discussion of simulations and assumptions.

\begin{table}[]
    \centering
    \begin{tabular}{|l|l|}
    \hline
        Decoherence source & Value \\
        \hline
        \hline
        Storage single-photon loss $(\kappa_1/\chi)$ & $q$ \\
        \hline
        Transmon loss $(\gamma_\downarrow/\chi)$ & $3q $ \\
        \hline
        Storage pure dephasing $(\kappa_\phi/\chi)$ & $0.01 q$ \\
        \hline
        Transmon pure dephasing $(\gamma_\phi/\chi)$ & $1.5q$ \\
        \hline
        Transmon heating $(\gamma_\uparrow/\chi)$ & $0.015q$ \\
        \hline
        
    \end{tabular}
    \caption{Single-parameter noise model used for gate simulations in the main text. All decoherence rates for the storage and transmon are varied in proportion with the parameter $q$, with $q\sim 10^{-5} - 10^{-4}$ achievable assuming current state-of-the-art coherence.}
    \label{tab:noise_model}
\end{table}

\subsection{Gate performance}

\begin{figure*}
    \centering
    \includegraphics[width=0.9\textwidth]{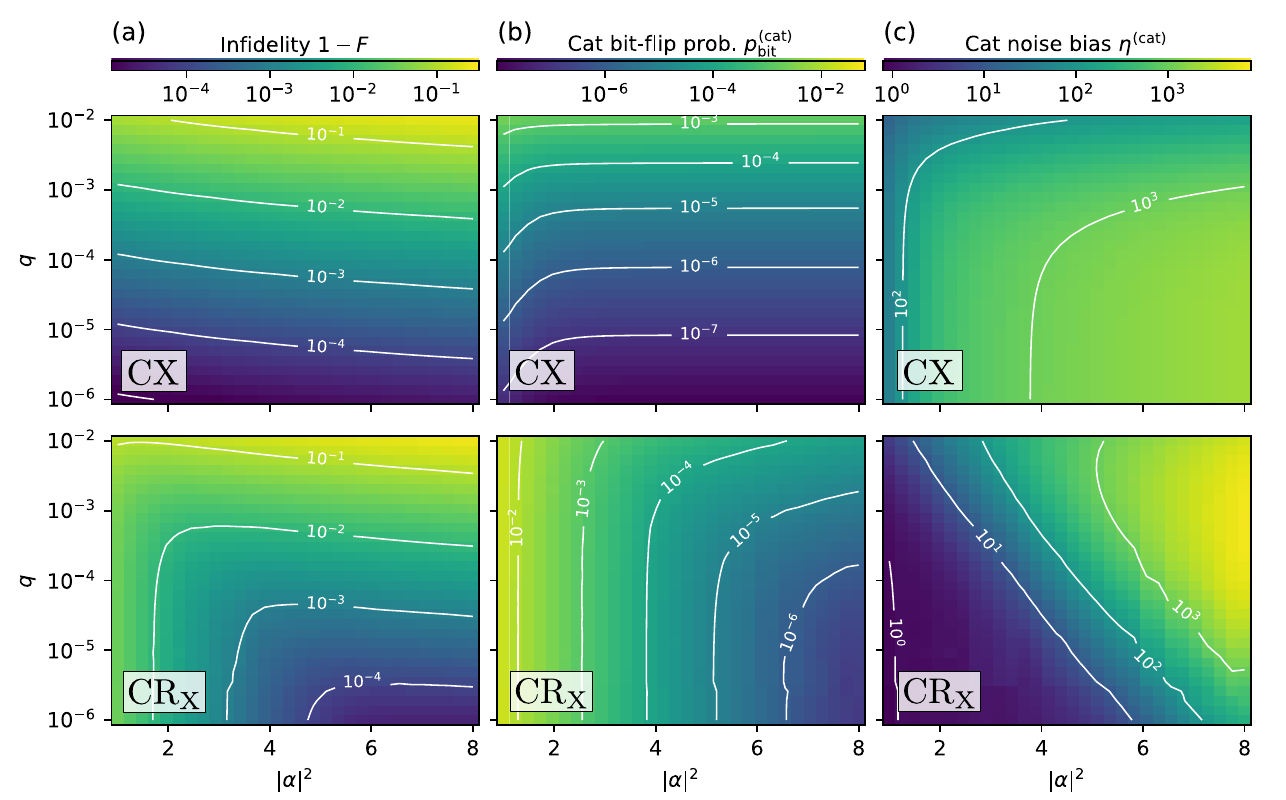}
    \caption{Error and bias of transmon-controlled \cX{} (top row) and cat-controlled \cRX{} (bottom row) gates.  Pauli error channels are extracted from master equation simulations of the gates, and the resultant \textbf{(a)} average gate infidelity, \textbf{(b)} cat bit-flip probability, and \textbf{(c)} cat noise bias are plotted as a function of $q$ (which specifies the decoherence rates of the cat and transmon per \cref{tab:noise_model}) and $|\alpha|^2$.  }
    \label{fig:cX_cRX_composite}
\end{figure*}

To quantify the gates' performance, we perform master equation simulations with decoherence rates as specified in \cref{tab:noise_model}. 
From these simulations we extract a Pauli error channel describing each gate's errors as discussed in \cref{app:extracting_error_channels}. We use these Pauli channels to compute the average gate infidelity $1-F$~\cite{nielsen2002simple}, cat bit-flip probability $p_\text{bit}^\text{(cat)}$, and the cat noise bias $\eta^\text{(cat)}$ as a function of $|\alpha|^2$ and $q$.

The average gate infidelity of the \cX{} and \cRX{} gates is shown in \cref{fig:cX_cRX_composite}(a). At sufficiently large $|\alpha|^2$ the gate infidelities of the two gates gates are similar. Both become dominated by cat Z errors, which occur with probability $ \kappa_1|\alpha|^2 (\pi/\chi) = \pi |\alpha|^2 q$. At smaller $|\alpha|^2$, the infidelities of the two gates display qualitatively different behavior. While the \cX{} gate is only limited by decoherence, the \cRX{} gate is also limited by coherent errors in the selective transmon pulses (see \cref{sec:pulse_shaping}). Low infidelities, $1-F<10^{-3}$, can be achieved for both gates simultaneously for realistically achievable values of $q$ in the range $10^{-5} - 10^{-4}$.

\cref{fig:cX_cRX_composite}(b) shows $p_\text{bit}^\text{(cat)}$, the probability of cat bit-flip error (i.e.~the probability of an $X$- or $Y$-type Pauli error on the cat), for each gate. The clear qualitative difference between $p_\text{bit}^\text{(cat)}$ between the \cX{} and \cRX{} gates highlights that the \cX{} is only moderately noise biased, while the \cRX{} is exponentially noise biased. In particular, for \cX{}, as $|\alpha|^2$ increases $p_\text{bit}^\text{(cat)}$ approaches the lower bound
\begin{equation}
\label{eq:transmon_bit_flip_limit}
    p_\text{bit, \cX{}}^\text{(cat)} \geq \frac{1}{2}\left[\frac{\gamma_\uparrow}{2}\frac{\pi}{\chi} + \left(\frac{\gamma_\downarrow}{2}\frac{\pi}{\chi}\right)^2\right],
\end{equation}
where the two terms correspond to the probabilities of cat bit-flips induced by transmon heating and double decay, respectively. Despite the lack of exponential suppression, this probability can be small because of the quadratic suppression of the transmon's dominant loss errors and small heating rates at low temperature.  
In contrast, for \cRX{}, $p_\text{bit}^\text{(cat)}$ continues to decrease exponentially with $|\alpha|^2$. This is because transmon errors cannot propagate through the \cRX{} to large phase space rotations of the storage mode; see \cref{App:cRX_details} for a detailed discussion.

Finally, \cref{fig:cX_cRX_composite}(c) shows the cat noise bias $\eta^\text{(cat)}$ for each gate. We define $\eta^\text{(cat)} \equiv p_Z^\text{(cat)}/p_\text{bit}^\text{(cat)}$, where $p_Z^\text{(cat)}$ is the probability of a $Z$-type Pauli error on the cat.
Importantly, while only the \cRX{} exhibits exponentially increasing bias with $|\alpha|^2$, both gates nevertheless exhibit large bias at modest $|\alpha|^2$. In particular, for realistic $q$, $\eta^\text{(cat)} > 10^3$ is simultaneously achievable for both gates. 
The combination large bias and low infidelity can enable significant hardware-efficiency improvements for error correction, as we show in the next section.


\section{Logical memory performance projections}
\label{Sec:logical}

By leveraging the \cX{} and \cRX{} gates, hybrid cat-transmon surface codes can be constructed that are capable of correcting both the cats' dominant $Z$ errors and their suppressed $X$ errors (see \cref{fig:architecture}). In this section, we simulate error correction with these codes to quantify their performance as logical memories. We first assess how realistically achievable error rates compare to the error threshold, then we quantify the hardware-efficiency benefit of the architecture's noise bias by comparing logical memory overhead to the case without biased noise. All simulations in this section employ circuit-level noise, with $d_Z$ rounds of noisy syndrome measurements simulated using Stim~\cite{gidney2021stim}, and minimum weight perfect matching decoding performed using PyMatching~\cite{higgott2022pymatching}.

\subsection{Error threshold}
\label{sec:error_threshold}

To determine the error threshold, we continue with the noise model in \cref{tab:noise_model} and directly use the \cX{} and \cRX{} Pauli error channels from \cref{Sec:achievable} in our circuit level noise model. In addition, we also specify error channels for cat idling and transmon readout and reset operations (see syndrome extraction circuits in \cref{fig:architecture}). We take 1\% combined error probability for transmon readout and reset~\footnote{The simulations assume 0.5\% error each for transmon readout and reset, but the combined error probability (1\%) is the only physically relevant parameter, since reset errors can be propagated to readout errors in the next round. }.
We assume that the total duration of the readout and reset operations is the same as the duration of the \cX{} and \cRX{} gates, i.e.~$\pi/\chi$, so that the noise model in \cref{tab:noise_model} can also be directly used in simulations of cat idling during transmon readout and reset. We then extract a Pauli error channel for cat idling subject to dissipative stabilization. We assume only a modest engineered dissipation rate relative to the dispersive coupling, $\kappa_2= \chi/10$, which is sufficient to prevent accumulation of leakage out of the cat code space (see \cref{app:pulsed_stabilization}).

\begin{figure}
    \centering
    \includegraphics[width=\columnwidth]{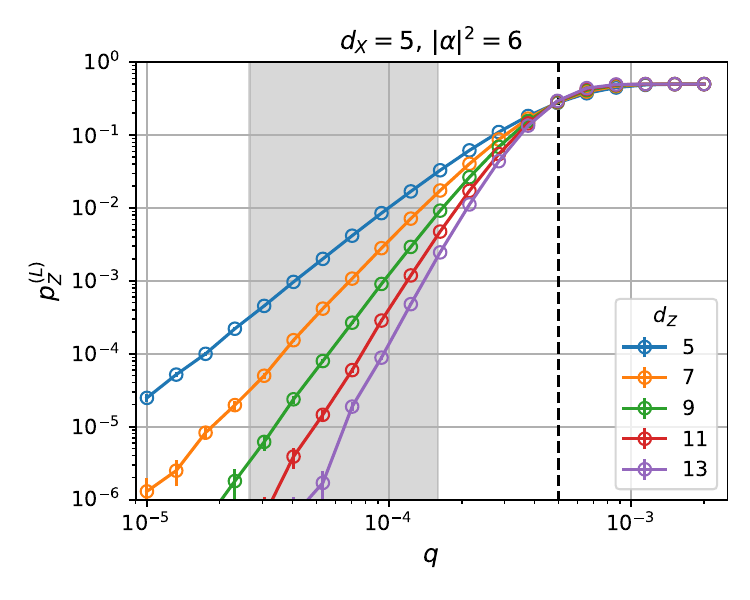}
    \caption{Logical $Z$ memory error rates (per $d_Z$ rounds) of the cat-transmon architecture. The logical error rate for different code distances is plotted as a function of $q$ at fixed $d_X=5$ and $|\alpha|^2=6$. These fixed values are sufficient to achieve $p_X^{(L)}\ll p_Z^{(L)}$ for all $q$ and $d_Z$ shown (see main text).  The ``threshold'' $q \approx 5\times 10^{-4}$ is indicated by the dashed line. Accessible $q$ values for current state-of-the-art devices are indicated by the gray region, which is defined here by computing the max and min achievable $q$ for $\kappa_1$ and $\chi$ in the ranges $\kappa_1^{-1} \in[1,2]\text{ms}$ and $\chi/2\pi\in[1,3]\text{MHz}$. $10^{7}$ samples are taken for each data point, and error bars indicate 95\% confidence intervals. }
    \label{fig:threshold}
\end{figure}

In \cref{fig:threshold}, we plot the logical $Z$ error probability after $d_Z$ rounds of noisy syndrome measurement, $p_Z^{(L)}$,  as a function of the $d_Z$ and $q$, at fixed $d_X = 5$ and $|\alpha|^2 = 6$. We observe exponential suppression of $p_Z^{(L)}$ with increasing $d_Z$ below a ``threshold'' value of $q = 5\times 10^{-4}$. To place this value in context, the region of realistically achievable $q$ for near-term state-of-the-art devices is shaded in gray. We see that error rates more than 10 times below threshold---the ``deep sub-threshold regime---are achievable in near-term devices, even with engineered dissipation of only modest strength.

It should be noted that we use use the term threshold loosely here. The logical $X$ error rate after $d_Z$ syndrome measurement rounds,  $p_X^{(L)}$, is lower bounded at fixed $d_X$ and $|\alpha|^2$, so identifying the proper asymptotic threshold would eventually also require increasing $d_X$. In practice, however, the code distances and error rates required to reach algorithmically relevant logical error rates are more relevant than the asymptotic threshold. Accordingly, $d_X = 5$ and $|\alpha|^2 = 6$ were chosen because they are sufficiently large to ensure that $p_X^{(L)}\ll p_Z^{(L)}$,  with $p_X^{(L)} < 10^{-10}$ in the deep sub-threshold regime (see \cref{App:logical_err_fitting}). Additionally,  we note that while XZZX surface code actually offers a higher threshold, we focus on the standard surface code here because it offers lower overheads in the deep sub-threshold regime (see \cref{App:code_comparison}).

\subsection{Comparison to unbiased noise}

\begin{figure}
    \centering
    \includegraphics[width=\columnwidth]{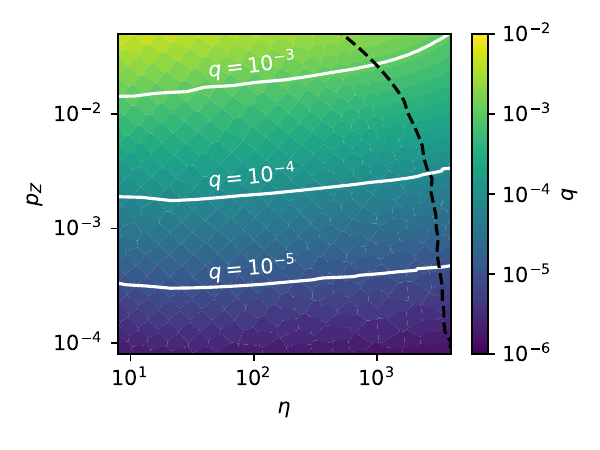}
    \caption{Achievable bias and $Z$ error probability. Colors indicate the maximum $q$ sufficient to achieve a given value of $p_Z$ (Z-type error probability) and $\eta$ (bias). To guide the eye, contours of constant $q$ are shown: any point in parameter space above a given contour is achievable with that value of $q$. The black dashed line indicates the maximum beneficial bias (i.e.~the point beyond which the bias only increases because of increasing cat $Z$ error probability without also decreasing $X$ error probability; see \cref{App:optimal_alpha} for details). The plot is produced by simulating \cX{} and \cRX{} gates at different values of $|\alpha|^2$ and $q$, then computing $p_Z$ and $\eta$ from their Pauli error channels as described in \cref{App:validating_simplified_model}. For visual clarity, Voronoi cells are constructed around these nonuniformly-distributed parameter space points, and each cell is colored according to the corresponding $q$.
}
    \label{fig:achievable_bias_pZ}
\end{figure}

\begin{figure*}
    \centering
    \includegraphics[width=\textwidth]{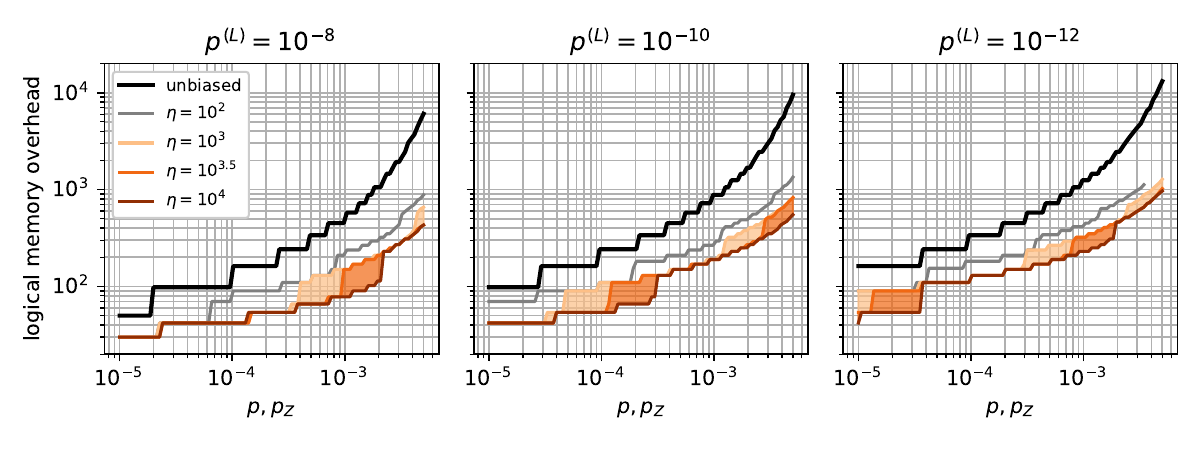}
    \caption{Logical memory overheads ($=2d_Xd_Z-1$) for different target $p^{(L)}$ per $d_Z$ rounds are computed for unbiased noise (black lines) and biased noise (gray, orange lines) as a function of the respective error parameters, $p$ and $p_Z$. Different biases in the achievable range $\eta \sim 10^{3}-10^{4}$ are indicated by different shades of orange. Regions between these curves are shaded to visually highlight how varying the bias over this range impacts the overhead. Overheads for lower bias, $\eta = 10^2$, are plotted in gray for reference. For the biased noise curves, small jumps in overhead correspond to increases in $d_Z$, while large jumps correspond to increases in $d_X$.
   Overheads are computed by extrapolating from fits as described in \cref{App:logical_err_fitting}. 
   }
    \label{fig:logical_overheads}
\end{figure*}

To more readily enable the comparison with the unbiased noise case, in this subsection we adopt a simplified circuit-level noise model to describe the cat-transmon architecture. In this simplified model, a biased Pauli error channel is applied after every two-qubit gate, and after data qubit idling during ancilla qubit readout. The biased error channels are specified in terms of two parameters, $p_Z$ and $\eta$: all $Z$-type errors have equal probability, with the total probability of a $Z$-type error given by $p_Z$, and the total probability of any other error is $p_Z/\eta$.  
This biased error model is compared with an unbiased error model where the same two-qubit gates and data idling operations are instead subject to a uniform depolarizing channel with entanglement infidelity~\cite{nielsen2010quantum} $1-F_e = p$.  The biased error channel's entanglement fidelity is given by $1-F_e = p_Z (1-1/\eta)$, which approaches $p_Z$ for large bias $\eta$, so that comparing $p$ and $p_Z$ is reasonable. Both models assume 1\% combined ancilla qubit readout and reset error probability.

We find that this simplified biased noise error model is sufficient to provide quantitatively accurate predictions for the the cat-transmon architecture's logical error probabilities. 
In particular, in \cref{App:validating_simplified_model} we show how the parameters $p_Z$ and $\eta$ can be extracted from the gates' full Pauli error channels, such that the logical errors with the simplified error model and with the full error model used in \cref{sec:error_threshold} agree well.
To show what values of $p_Z$ and $\eta$ are achievable in the cat-transmon architecture, in \cref{fig:achievable_bias_pZ} we plot the achievable values of these parameters as a function of $q$. Consistent with the findings in \cref{Sec:achievable}, $p_Z \sim 10^{-3}$ and $\eta \sim 10^{3.5}$ are simultaneously achievable with state-of-the-art coherence, i.e.~$q$ between $10^{-4}$ and $10^{-5}$. In \cref{app:alternative_noise_models}, we present analogous plots of the achievable $p_Z$ and $\eta$ for alternative noise models. There we find that slightly larger bias $\eta \sim 10^{4}$ is achievable when the storage and transmon lifetimes are comparable. In the analysis below, we thus consider the range $\eta \sim 10^{3} - 10^{4}$ as broadly representative of what is achievable with the cat-transmon architecture.

The metric we use to compare the biased and unbiased noise cases is the logical memory overhead, defined as the minimum number of qubits (data and ancilla) required to reach a given target total logical error $p^{(L)} = p^{(L)}_X + p^{(L)}_Z$. For a rectangular surface code with $X$ and $Z$ distances $d_X, d_Z$, the total number of qubits is given by $2 d_X d_Z-1$. To compute the overhead, we thus find the values $d_X$ and $d_Z$ that minimize $2 d_X d_Z-1$ subject to the constraint that logical error falls below the given $p^{(L)}$. In the unbiased noise case, square codes $d_X = d_Z$ are optimal, but in the biased noise case rectangular codes $d_X < d_Z$ can offer lower overheads (see \cref{App:logical_err_fitting}).

In \cref{fig:logical_overheads}, 
we plot logical memory overheads for the unbiased and biased error models as a function of $p$ and $p_Z$, respectively, for different choices of the bias $\eta$ and target logical error $p^{(L)}$. Even for the modest bias achievable with the cat transmon architecture, $\eta \sim 10^{3}-10^{4}$, there is significant reduction in the overhead relative to the unbiased noise case. Indeed, in the deep sub-threshold regime, we find that realistically achievable noise biases can provide up to a factor 5 reduction in logical memory overhead relative to the case without noise bias. For example, with $\eta = 10^{3.5}$ and $p_Z=10^{-3}$ for the biased noise case, the overhead to achieve $p^{(L)} = 10^{-10}$ is $\sim 200$ qubits, while the overhead to achieve the same $p^{(L)}$ with $p = 10^{-3}$ is nearly $\sim 1000$ qubits for the unbiased case. 

With \cref{fig:logical_overheads}, one can also compare the physical error required to achieve the target logical error at a given overhead. Consider again the values $p_Z\sim 10^{-3}$ and $\eta \sim 10^3 - 10^{4}$ that are achievable with the cat-transmon architecture assuming state-of-the-art coherence. Across the different $p^{(L)}$, we see that the overhead of the biased-noise architecture at these $p_Z, \eta$ values is comparable to the overhead of the unbiased-noise architecture with $p \sim 10^{-5}-10^{-4}$. That is, the logical memory performance of the cat-transmon architecture with realistically achievable physical error and bias is roughly equivalent to that of a generic unibiased noise architecture with physical error in the range $10^{-5}-10^{-4}$. 

To provide context for this comparison, we consider a realization of an unbiased noise architecture with transmon qubits and crudely estimate the transmon lifetimes that would be required to achieve $p \sim 10^{-5}-10^{-4}$. Our unbiased error model assumes two-qubit gates and data qubit idling during readout each have error $p$, so that the average probability of a data qubit error is roughly $5p$ per error correction cycle. Assuming an error correction cycle duration of $T_\text{cycle} = 1\,\mu s$, and that the performance of transmon-transmon gates is limited only by decoherence, the range $p \sim 10^{-5}-10^{-4}$ thus corresponds to transmon lifetimes in the range of $T_\text{cycle}/5p = 2-20$ ms. Current state-of-the-art transmon lifetimes lie in the range $\sim 0.3-1$ ms~\cite{place2021new,wang2021transmon,tuokkola2024methods}, approaching the minimum lifetime required for $p=10^{-4}$ but below that required for $p=10^{-5}$. That said, we emphasize that this is only a crude estimate, and indeed making a more accurate comparison would require additional considerations beyond coherence limits. For example, the cats' dissipative stabilization also serves as built-in means of leakage reduction, so one should also account for any additional cost associated with leakage reduction in the unbiased case~\cite{ghosh2015leakage, miao2023overcoming, yang2024coupler}. An accurate comparison will also require accounting for integration challenges (and opportunities) that arise when building larger devices. Identifying and quantifying these integration aspects for the cat-transmon architecture is thus an important direction for future research.


\section{Discussion and Outlook}
\label{sec:discussion}

In this work, we have proposed and analyzed the hybrid cat-transmon architecture---a scalable architecture for hardware-efficient error correction. In this archtiecture, the dissipative cats' noise bias enables hardware efficiency, and the use of ancillary transmons enables syndrome extraction with simple, high-fidelity cat-transmon entangling gates. To correct the cats' dominant $Z$ errors, we employ a moderately noise-biased \cX{} gate that relies only on free evolution under the native dispersive coupling---a highly simple implementation that has enabled multiple experimental demonstrations~\cite{rosenblum2018fault,reinhold2020error,repcat}. To correct the cats' suppressed $X$ errors and hence enable full scalability, we propose the exponentially noise-biased \cRX{} gate that relies primarily on simple number-selective pulses, again enabled by the native dispersive coupling. 
In comparison to cat-cat gates, these cat-transmon gates can be fast ($\pi/\chi \sim 200$ ns) and high fidelity ($F\sim 99.9\%$), assuming MHz-scale dispersive couplings and ms-scale storage $T_1$. They also achieve sizable noise bias ($\eta \sim 10^{3}-10^{4}$), and though the noise bias of this architecture is ultimately limited by transmon errors, we find that significant hardware efficiency improvements are possible. For realistically achievable noise bias and error rates, projected logical memory overheads for the cat-transmon architecture are comparable to those of a generic unbiased noise architecture with physical error probability in the range $\sim10^{-5}- 10^{-4}$. 

The performance of the cat-transmon architecture is largely decoherence limited, so that the ratios of the cat and transmon's decoherence rates relative to the dispersive coupling strength are the main figures of merit that dictate logical performance.
On one hand, in 3D devices, ratios of the storage mode loss rate to dispersive coupling, $\kappa_1/\chi =q$, in the range $q\sim 10^{-5}-10^{-4}$ are realistically achievable assuming MHz-scale dispersive couplings and ms-scale storage $T_1$, which would place such devices in the deep subthreshold regime. Recent work~\cite{milul2023superconducting} demonstrating ultrahigh storage mode lifetimes of 10s of milliseconds suggests further room for improvement. However, increasing storage lifetime by itself without also increasing transmon lifetime has diminishing returns for the performance of the cat-transmon architecture. In particular, cat $Z$ error rates are dictated by $T_1$ of dressed storage modes (i.e.~storage modes hybridized with transmons), rather than the intrinsic $T_1$ of isolated storage modes, and the transmon lifetime ultimately limits the dressed storage $T_1$ via the inverse Purcell effect. Moreover, transmon errors can directly propagate to cat $Z$ errors through the \cRX{} gate, such that transmon coherence limits the achievable $p_Z$ (see~\cref{app:alternative_noise_models}). While currently achievable transmon lifetimes on the order of $\sim 100\, \mu s$ already suffice to access the deep sub-threshold regime with ms-scale storage lifetimes, improvements in transmon lifetime would thus be required for this architecture to fully capitalize on ultra-high intrinsic storage lifetimes.

On the other hand, 2D devices are naturally compact and scalable, but further improvements in coherence are required to push them into the deep subthreshold regime. 
Quantum acoustic devices have the potential to offer both high coherence and high scalability~\cite{maccabe2020nano,arrangoiz2019resolving}, but preserving the high coherence in integrated systems remains an outstanding challenge.
Independent of the storage modes' physical implementation, efforts to increase the dispersive coupling strength~\cite{ye2021engineering} also provide a promising avenue to improved performance. A key challenge in pushing beyond MHz-scale dispersive couplings will be suppressing the presence or mitigating the impacts of spurious nonlinearities.  

While we have focused on the specific case of a transmon ancilla, we emphasize that the entangling gates we employ apply equally well to any type of dispersively coupled ancilla qubit, and alternative choices of the ancilla qubit could potentially enable further performance improvements. One intriguing possibility in this regard would be to substitute the transmon for a fluxonium. Fluxonium qubits are appealing generally because of their long coherence times and large anharmonicites~\cite{nguyen2019high,bao2022fluxonium,zhang2021universal}. In the context of this architecture, the ability to engineer a large fluxonium noise bias by operating off sweet spot~\cite{lin2018demonstration}, and to engineer large dispersive couplings thanks to the large anharmonicity are appealing properties. In particular, for a fluxonium ancilla to offer an improvement over a transmon ancilla, the fluxonium-induced cat bit-flip probability during a \cX{}, $ p_\text{bit, \cX{}}^\text{(cat)}= \frac{1}{4}(\gamma_\uparrow+\gamma_\downarrow)/\chi$, should be smaller than the corresponding probability for the transmon ancilla, cf.~\cref{eq:transmon_bit_flip_limit}.

Modifications to either the \cX{} or \cRX{} gates offer another promising avenue for further improvement of the architecture. Indeed, these two gates are somewhat mismatched; where the \cX{} gate is only moderately noise biased yet capable of achieving high fidelity at small $|\alpha|^2$, the \cRX{} gate is exponentially noise biased but limited by coherent error at small $|\alpha|^2$. Instead, one can consider variations of the architecture where the benefits and drawbacks of the entangling gates are in better alignment. For example, one could retain the current \cX{} and modify the \cRX{} to allow for better small-$|\alpha|^2$ performance by sacrificing its exponentially noise-biased nature. This modification could be achieved, for instance, simply by extending the \cRX{} gate time beyond $\pi/\chi$, thereby mitigating coherent error at the price of more transmon-induced cat $X$ errors. The resultant small-$|\alpha|^2$ architecture would likely enjoy even higher thresholds against photon loss at the price of  reduced noise bias. Alternatively, one could retain the current \cRX{} and replace the \cX{} with an exponentially noise-biased gate. For example, one could employ a cat ancilla and exponentially noise-biased cat-cat gates for $X$-type syndrome measurements, while continuing to use the transmon ancilla and cat-transmon \cRX{} gate for $Z$-type syndrome measurements. This approach has the advantage of enabling the optimizations proposed in Ref.~\cite{le2023high} that mitigate the impact of the unfavorable $\sqrt{\kappa_1/\kappa_2}$ infidelity scaling of cat-cat entangling gates only in the context of $X$-type syndrome measurement. Combining this with \cRX{}-enabled $Z$-type syndrome measurements means one need not rely exclusively on the cats' $X$ error suppression to reach algorithmically-relevant logical error rates.

Finally, while in this work we have focused on logical memory performance of the surface code as the main figure of merit, the performance of fault-tolerant logical operations will need to be evaluated to provide a complete picture of the architecture's potential. In the cat-transmon architecture, fault-tolerant logical operations can be implemented via lattice surgery~\cite{horsman2012surface}. Because syndrome measurement is the basic primitive involved in lattice surgery, the use of cat-cat entangling gates can still be avoided in favor of syndrome measurement with ancillary transmons, so that the practical benefits of the cat-transmon architecture persist at the level of logical operations. While we leave detailed study of fault-tolerant operations via lattice surgery in the cat-transmon architecture to future work, we note that the underlying physical noise bias has the potential to offer even further improvement in the context of logical operations. For example, this noise bias can be exploited in the context of magic state distillation to reduce overheads further~\cite{singh2022high}. Furthermore, alternative code constructions that exploit the noise bias could be explored. As an example, one could consider a hierarchical construction~\cite{pattison2023hierarchical} where some bias is retained at the level of encoded surface code logical qubits, which are then concatenated into a quantum low-density parity-check (LDPC) code~\cite{breuckmann2021quantum, panteleev2022asymptotically,leverrier2022quantum,bravyi2024high} (using lattice surgery to enable any requisite long-range interactions). Further work on LDPC codes tailored to exploit the noise bias~\cite{ruiz2024ldpc,roffe2023bias} is needed evaluate the potential of this approach.

\section{Acknowledgements}

We thank Arne L. Grimsmo, Gregory S.~MacCabe, Shahriar Aghaeimeibodi, Yufeng Ye, Catherine Leroux, and Aleksander Kubica for useful discussions, as well as Colm A.~Ryan for helpful feedback on the manuscript. We also thank Simone Severini, Bill Vass, James Hamilton, Nafea Bshara, and Peter DeSantis at AWS, for their involvement and support of the research activities at the AWS Center for Quantum Computing.

\appendix

\section{Details of the \cRX{} gate}
\label{App:cRX_details}
In this section, we discuss certain features of the \cRX{} gate in further detail. First, we show that the pulse sequence presented in the main text indeed implements the \cRX{} operation. Next,  we demonstrate that cat bit-flip errors induced by transmon errors are exponentially suppressed with increasing $|\alpha|^2$. This allows us to identify the optimal choice of $|\alpha|^2$ for the cat-transmon architecture by weighing this exponential suppression against the bit-flip error saturation of the $\cX{}$ gate. Next, we derive the expressions given in the main text for the storage dephasing rate in the presence of the storage echo, and describe the generalized DRAG pulse shaping used to improve performance. Finally, we discuss additional applications of the \cRX{} gate that may be of independent interest. In particular we show how the \cRX{} gate can be used to implement fast, high-fidelity cat $Z$ and CZ rotations in a exponentially noise-biased manner, as well as single-shot cat $Z$ basis readout with readout error exponentially suppressed in $|\alpha|^2$. 

\subsection{Unpacking the pulse sequence}
In this section, we show that the pulse sequence presented in \cref{Sec:cRX} implements the \cRX{} unitary \cref{eq:cRX}. We begin by first showing that this unitary can also be implemented by a simplified sequence that omits the storage echo:
\begin{enumerate}
    \itemsep0em 
    \item Storage displacement: $D(+\alpha)$
    \item Selective transmon pulse: $R_X(\pi/2)$
    \item Unselective transmon pulse: $R_X(\pi)$
    \item Selective transmon pulse: $R_X(\pi/2)$
    \item Storage displacement: $D(-\alpha e^{i\phi})$
\end{enumerate}
Here $\phi = \chi T_\mathrm{sel.}$, where $T_\mathrm{sel.}$ is the selective pulse duration. This simplified sequence may be more convenient than the full \cRX{} sequence in situations where the storage echo is not needed or not helpful (e.g., if storage dephasing is negligible or is dominated by high-frequency components not mitigated by the echo).

Consider first the case where the cat qubit is initially in the state $\ket{-\alpha}$ prior to the gate. The displacement in step 1 enacts $\ket{-\alpha}\to\ket{0}$. Since the selective pulses act nontrivially when the storage mode is in vacuum, a net rotation $R_X(2\pi)$ is applied to the transmon in steps 2-4. The final displacement enacts $\ket{0}\to\ket{-\alpha e^{i\phi}}$. Next, consider the initial state $\ket{\alpha}$. The displacement in step 1 enacts $\ket{\alpha} \to \ket{2\alpha}$. As a result, only the unselective pulse in step 3 acts nontrivially, and a net rotation of $R_X(\pi)$ is applied to the transmon in steps 2-4. During these steps, the transmon-storage dispersive coupling causes storage mode state rotate $\ket{2\alpha}\to \ket{2\alpha e^{i\phi}}$. Note that this rotation is deterministic: the unselective transmon pulse ensures that, regardless of its initial state, the transmon is excited for time $T_\mathrm{sel.}$ during these steps, so that the dispersive phase accumulated is $\chi T_\mathrm{sel.} = \phi$. The final displacement enacts $\ket{2\alpha e^{i\phi}} \to \ket{\alpha e^{i\phi}}$. The net effect is thus to implement the operation
\begin{equation}
    \ket{\alpha e^{i \phi}}\bra{\alpha} \otimes R_X(\pi) + \ket{-\alpha e^{i\phi}}\bra{-\alpha} \otimes R_X(2\pi). \nonumber
\end{equation}
Up to a local operations (a rotation of the storage mode phase by $-\phi$ and a rotation of $R_X(-\pi)$ on the transmon), this operation is equivalent to the \cRX{} unitary in \cref{eq:cRX}.

Now, we consider the full \cRX{} sequence:
\begin{enumerate}
    \itemsep0em 
    \item Storage displacement: $D(+\alpha)$
    \item Selective transmon pulse: $R_X(\pi/4)$
    \item Unselective transmon pulse: $R_X(\pi)$
    \item Selective transmon pulse: $R_X(\pi/4)$
    \item Storage displacement: $D(-2\alpha e^{i\phi})$
    \begin{enumerate}[]
        \item Storage displacement: $D(-\alpha e^{i\phi})$ 
        \item Storage displacement: $D(-\alpha e^{i\phi})$
    \end{enumerate}
    \item Selective transmon pulse: $R_X(-\pi/4)$
    \item Unselective transmon pulse: $R_X(-\pi)$
    \item Selective transmon pulse: $R_X(-\pi/4)$
    \item Storage displacement: $D(\alpha e^{2i\phi})$
\end{enumerate}
To aid in the explanation of this sequence, we have divided the storage displacement $D(-2\alpha e^{i\phi})$ in step 5 into two separate displacements $D(-\alpha e^{i\phi})$ in steps 5(a) and 5(b). With this division, this sequence can be roughly understood as two repetitions of the simplified sequence above. By an analogous argument, then, steps 1-5(a) implement
\begin{equation}
    \ket{\alpha e^{i \phi}}\bra{\alpha} \otimes R_X(\pi)+\ket{-\alpha e^{i\phi}}\bra{-\alpha} \otimes R_X(3\pi/2), \nonumber
\end{equation}
and steps 5(b)-9 implement
\begin{equation}
    \ket{\alpha e^{2i \phi}}\bra{\alpha e^{i\phi}} \otimes R_X(-3\pi/2)+ \ket{-\alpha e^{2i\phi}}\bra{-\alpha e^{i\phi}} \otimes R_X(-\pi). \nonumber
\end{equation}
Applied in sequence, the net operation is,
\begin{equation}
    \ket{\alpha e^{2i \phi}} \bra{\alpha} \otimes R_X(-\pi/2) + \ket{-\alpha e^{2i\phi}}\bra{-\alpha} \otimes R_X(\pi/2). \nonumber
\end{equation}
Up to local operations (a rotation of the storage mode phase by $-2\phi$ and a rotation of $R_X(\pi/2)$ on the transmon), this operation is equivalent to the \cRX{} unitary in \cref{eq:cRX}.

\subsection{Exponential suppression of cat bit flips and optimal $|\alpha|^2$}
\label{App:optimal_alpha}

Like in the \cX{} gate, the transmon-storage dispersive coupling allows transmon errors during the \cRX{} gate to propagate to spurious rotations of the storage mode, which can cause cat $X$ errors. However, so long as these spurious storage mode rotations $\delta$ are sufficiently small, i.e.~$\delta<\pi/2$, the probability of an $X$ error will be exponentially suppressed with $|\alpha|^2$. In what follows we demonstrate that so long as the \cRX{} gate time is sufficiently short, roughly $T_\mathrm{cR_X} \lesssim \pi/\chi$, the amount of spurious rotation is sufficiently small to achieve this exponential suppression. 

The amount of storage mode rotation induced by transmon errors can be determined from the \cRX{} pulse sequence. To calculate the worst-case spurious phase accumulation due to transmon errors, consider the example of a \cRX{} gate applied to the the initial state $|+\alpha\rangle$. During the first half of the gate, the largest spurious rotations of the storage mode are induced by transmon bit-flip errors ($|g\rangle \leftrightarrow |e\rangle$) immediately before or after the unselective pulse of step 3. Such an error effectively cancels the action of the unselective pulse, so that the transmon remains in the same state during at least the entire first half of the \cRX{} pulse sequence. In this situation, the transmon-dispersive coupling thus results in a storage mode phase accumulation of either $0$ or $2\phi$, depending on the transmon state, which differ by $\pm \phi$ from the expected phase accumulation of $\phi$ during these steps. 
 
Continuing with this example, the subsequent displacement in step 5 no longer maps the storage mode to the vacuum state, so the storage continues to accumulate spurious phase during the second half of the gate. Crudely speaking, we expect that yet another transmon error immediately before or after the unselective pulse in step 7 could result in additional spurious storage phase accumulation of $\approx \phi$ by the same logic as above. The total spurious phase accumulation over the full gate, $\delta$, would thus be $\delta \approx 2\phi$. Requiring $T_\mathrm{sel.}\lesssim \pi/4\chi$ (equivalent to $T_\mathrm{CR_X}\lesssim \pi/\chi$ when the selective pulses dominate the total gate duration) thus ensures that $\delta \lesssim \pi/2$, so that bit-flip errors would be exponentially suppressed.

The above requirement is only approximate since we have neglected the fact that the spurious rotations and \cRX{} displacements do not commute. One can account for this the spurious phase accumulation more precisely as follows. Consider the specific instance of the above example where the transmon errors are such that the transmon remains in $\ket{g}$ throughout the whole gate. In this case, the only nontrivial operations are the three displacements, which combine to map initial state $\ket{\alpha}$ to $\ket{\beta}$, where $\beta = \alpha(2-2e^{i\phi}+e^{2i\phi})$. Relative to the expected final state of $\ket{\alpha e^{2i\phi}}$, the final state $\ket{\beta}$ corresponds to a spurious rotation of
\begin{equation}
    \delta = \arctan\left[\frac{2(\sin\phi-\sin2\phi)}{1-2\cos\phi+2\cos2\phi}\right].
\end{equation}
Note that to leading order in $\phi$, we have $|\delta| \approx 2\phi$, consistent with the simplified argument above. Using the more precise expression, we find that $|\delta| < \pi/2$ for $\phi < \pi/5$, corresponding to the requirement that $T_\mathrm{sel} \leq \pi/5\chi$ for exponential suppression of cat bit-flip errors.

\begin{figure}
    \centering
    \includegraphics[width=\columnwidth]{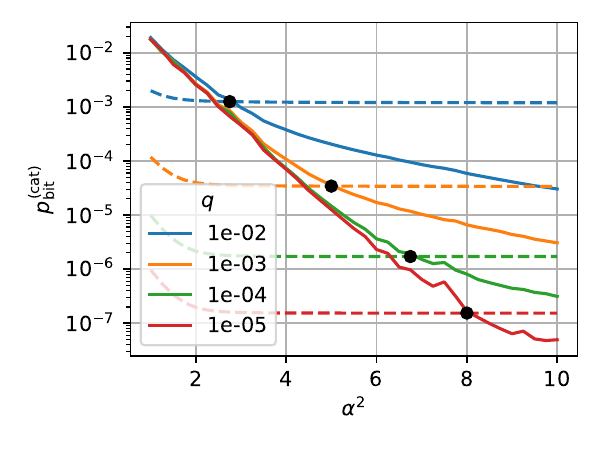}
    \caption{Cat bit-flip probability of \cRX{} (solid) and \cX{} (dashed) gates. For a given $q$, the sum of the \cX{} and \cRX{} bit-flip probabilities initially decreases exponentially with $|\alpha|^2$, since \cRX{} bit-flips are suppressed exponentially with $|\alpha|^2$. The sum will saturate once a sufficiently large $\alpha$ (black dots) is reached, owing to \cX{} bit-flips induced by transmon double-decay and heating. Operating at $|\alpha|^2$ larger than the black dots is thus not beneficial. }
    \label{fig:cXcRX_bit_flip_probs}
\end{figure}

We illustrate this exponential bit-flip suppression numerically in \cref{fig:cXcRX_bit_flip_probs}, where the simulated cat bit-flip probability of the \cRX{} gate is shown (solid lines) as a function of $|\alpha|^2$ for different choices of $q$. As in the main text, we have chosen pulse durations so that $T_\mathrm{CR_X}=\pi/\chi$ (and in particular $T_\mathrm{sel.}= \pi/5\chi$), so that we expect exponential suppression with increasing $|\alpha|^2$. 

The desired exponential suppression is indeed observed in \cref{fig:cXcRX_bit_flip_probs}, with different suppression exponents at different values of $|\alpha|^2$. This behavior is explained by the fact that transmon errors can be induced by different mechanisms, namely decoherence and selective pulse errors. At small $|\alpha|^2$, transmon errors are dominated by selective pulse errors, which are themselves suppressed with $|\alpha|^2$. As a result, in this regime we see a rapid exponential suppression, since both the probability of the selective pulse errors and the cat phase-space separation are suppressed with $|\alpha|^2$. At sufficiently large $|\alpha|^2$, the suppression exponent changes, as the transmon errors begin to be dominated by decoherence (controlled by the parameter $q$). The suppression exponent is smaller because, unlike selective pulse errors, the transmon decoherence errors are not themselves suppressed with $|\alpha|^2$. At lower transmon decoherence rates (smaller $q$), the rapid exponential suppression associated with selective pulse errors persists for longer before the more moderate suppression associated with transmon decoherence takes over. 

Finally, for the sake of comparison, in \cref{fig:cXcRX_bit_flip_probs} we also plot the cat bit-flip probability of the \cX{} gate (dashed lines) at the same values of $q$. As discussed in the main text, the \cX{} bit-flip probabilities saturate to a limit dictated by transmon double decay and heating errors (see \cref{eq:transmon_bit_flip_limit}). In practice, there are diminishing returns to operating at a value of $|\alpha|^2$ larger than that where the the \cRX{} and \cX{} bit-flips intersect (black dots). Beyond this point, the total cat bit-flip probability during error correction remains essentially constant, limited by the \cX{} gate, while the cat phase-flip probability increases linearly. Moving beyond this point would still increase the noise bias, but since the physical error probabilities are only increasing, there would be no accompanying improvement in logical performance. These intersection points therefore specify the maximum beneficial bias achievable in the cat-transmon architecture, and they are indicated as the black dashed line in \cref{fig:achievable_bias_pZ} of the main text.

\subsection{Analytical expressions for storage dephasing}
\label{App:storage_echo}
In this section, we derive the expression for the storage dephasing rate, \cref{eq:dephasing_rate} of the main text, and apply it to calculate the the dephasing rate for different types of noise with and without the storage echo. 

\cref{eq:dephasing_rate} follows directly from Eq.~(18) of Ref.~\cite{cywinski2008enhance}, with the modification that the dephasing rate is enhanced in case of the displaced cat in the \cRX{} gate, owing to the larger photon number of the displaced coherent state $|2\alpha\rangle$. In particular, the enhancement factor is given by the squared mean photon number of this state $(\braket{2\alpha|a^\dagger a|2\alpha})^2 = 16|\alpha|^4$. Multiplying the result of Ref.~\cite{cywinski2008enhance} by this enhancement factor yields \cref{eq:dephasing_rate}, which we repeat here for convenience:
\begin{equation}
    \Gamma_\phi(t) = 16 |\alpha|^4 \int_0^\infty \frac{d\omega}{2\pi} S(\omega) \frac{F(\omega t)}{\omega^2}.
\end{equation}

We proceed to use this expression to compute the dephasing rate $\Gamma_\phi(t)$ for different dephasing noise spectra, comparing the cases with and without the storage echo. Without an echo, the filter function is given by $F(\omega t) = 2\sin^2(\omega t/2)$, and with the echo $F(\omega t) = 8 \sin^4(\omega_t/4)$~\cite{cywinski2008enhance}. We begin with the case of white noise, defined by a constant noise spectrum $S(\omega)=A^2$, for some constant $A$. Regardless of whether the echo is applied, we obtain 
\begin{equation}
    \Gamma_\phi^\text{(white)}(t) = 16|\alpha|^4 A^2 t/2.
\end{equation}
The case of $1/f$ noise, or pink noise, is defined by $S(\omega) = 2\pi A^2/\omega$. Without the echo, this yields
\begin{equation}
    \Gamma_\phi^\text{($1/f$, no echo)}(t) = 16|\alpha|^4 (At)^2 \log\frac{e^{3/2}-\gamma}{\omega_\text{ir} t},
\end{equation}
where $\gamma$ is the Euler constant and $\omega_\text{ir}$ is the infrared cutoff (see, e.g., Ref.~\cite{didier2019ac}). 
With the echo, we obtain a different result, 
\begin{equation}
    \Gamma_\phi^\text{($1/f$, echo)}(t) = 16|\alpha|^4 (At)^2 \log2.
\end{equation}
At time scales that are short relative to the cutoff, $\omega_\text{ir} t \ll 1$, the echo can thus provide a significant reduction in the dephasing rate, though the magnitude of this improvement depends on the precise value of the cutoff. 

These expression are plotted alongside corresponding numerical results in \cref{fig:storage_echo}, showing good agreement. In the simulations, we have chosen values for the constants $A$ such that the $\Gamma_\phi(t=1/\kappa_\phi) = -16|\alpha|^2$. This way, for both white and $1/f$ noise, the time it takes for the single-photon coherence, $\ket{n=0}\bra{n=1}$, to decay by a factor of $1/e$ is the same, and this time defines the decay rate $\kappa_\phi$. To simulate $1/f$ noise, we perform trajectory simulations following the approach of Ref.~\cite{didier2019ac}. In particular, for each choice of noise strength, the dephasing rate is computed from $10^4$ individual trajectories, each of length $T$ and downsampled from a longer single trajectory, such that the infrared cutoff is given by $\omega_\text{ir} = 2\pi/(10^4 T)$.

\subsection{Pulse shaping}

\begin{figure}
    \centering
    \includegraphics[width=1\linewidth]{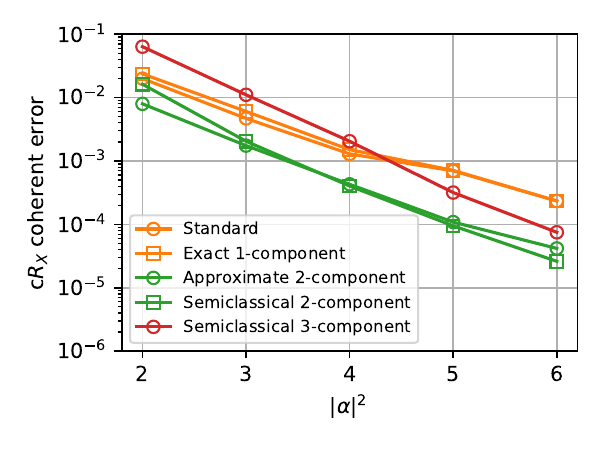}
    \caption{Comparison of \cRX{} error with different selective pulse shapes. Lines are colored according to whether they are designed to suppress 1 (orange), 2 (green), or 3 (red) undesired transitions.  Here, for numerical expediency, the \cRX{} coherent error is computed by averaging the infidelity over a small handful of initial cat-transmon states. In contrast, in \cref{fig:DRAG} the error is computed from the full \cRX{} Pauli error channel, hence the values in this figure differ slightly from the corresponding ones in \cref{fig:DRAG}.}
    \label{fig:sel_pulse_breakdown}
\end{figure}

In this section we describe the shaped pulses used to improve the performance of the \cRX{} gate. We consider four different types of selective pulses: \texttt{standard DRAG} pulses, \texttt{exact 1-component DRAG} pulses, \texttt{approximate 2-component DRAG} pulses, and \texttt{semiclassical 2-component DRAG} pulses. We define each pulse type in turn below, and the \cRX{} coherent error each pulse is plotted in \cref{fig:sel_pulse_breakdown}.  These pulses shapes are all derived directly following the approach in Ref.~\cite{motzoi2013improving}.  For numerical simulations of \cRX{} gates the in the main text, we optimize over these four pulse types. 

We consider a transmon drive of the form $H_\text{drive} = \Omega(t)/2\ket{e}\bra{g}) +\text{H.c.}$. The validity of this two-level approximation is discussed in \cref{App:sim_assumptions}. For our baseline pulse shapes, we consider a family of truncated Gaussian pulses~\cite{xu2022engineering}
\begin{equation}
     \Omega_0(t) = \exp\left(\frac{(t-T_\text{sel.}/2)^2}{2\sigma^2}\right) - \exp\left(\frac{(T_\text{sel.}/2)^2}{2\sigma^2}\right)
\end{equation}
where $T_\text{sel}.$ denotes the selective pulse duration, we fix $\sigma = T_\text{sel.}$, and the first $m-1$ derivatives of the pulse vanish at $t=0,T_\text{sel.}$.

Given a base pulse shape $\Omega_0$ (e.g.~a simple Gaussian pulse) we define the \texttt{standard DRAG} pulses as
\begin{equation}
    \Omega(t) = \Omega_0(t) + i p_1 \dot \Omega_0(t),
\end{equation}
along with a time-dependent pulse detuning
\begin{equation}
    \delta(t) = p_2 \Omega^2(t) + p_3,
\end{equation}
where for $\Omega_0(t)$ we take the $m=1$ truncated Gaussian pulse. 
The parameters $p_i$ are numerically optimized to maximize the performance of the pulse (see, e.g., Refs.~\cite{gambetta2011analytic,nguyen2022scalable}).

The \texttt{exact 1-component DRAG} pulse exactly implements a given rotation at a desired transition without inducing any rotation at a single undesired transition detuned by $\Delta_1$. That is, consider a simplified \cRX{}-like scenario where the storage mode state is either vacuum $\ket{0}$ or a single Fock state $\ket{n}$ (rather than the coherent state $\ket{2\alpha}$). In this simplified scenario, the goal is to drive the transmon nontrivially when the storage is in $\ket{0}$, and to not drive the transmon when the storage is $\ket{n}$. We thus have one desired transition at frequency $\omega_q$, and one undesired transition at $\omega_q + \Delta_1$, where $\Delta_1 = n\chi$. This simplified problem is solved exactly (i.e.~a perfectly selective drive is implemented) by the pulse shape 
\begin{equation}
\label{app_eq:e1_omega}
    \Omega(t) = \Omega_0(t) +i \frac{\Omega_0(t)\Delta(t)-\dot\Delta(t)\dot\Omega_0(t)}{\Omega_0(t)^2+\Delta(t)}
\end{equation}
with the time-dependent pulse detuning
\begin{equation}
\label{app_eq:e1_delta}
    \delta(t) =- \Re\Omega(t)\tan\left[\int_0^tdt' \Im\Omega(dt')\right],
\end{equation}
implementing a rotation by angle 
\begin{equation}
\label{app_eq:e1_theta}
    \theta=\int_0^T dt\,\Re\Omega(t)\sec\left[\int_0^{t} dt' \Im \Omega(t)\right].
\end{equation}
See Eq.~(5.4) in Ref.~\cite{motzoi2013improving} (which we note drops the $\propto\dot\Delta(t)$ term in the numerator).
In the above, $\Delta(t) = \delta(t) +\Delta_1$, and for $\Omega_0(t)$ we take the $m=1$ truncated Gaussian pulse. In the context of the \cRX{} gate, this pulse shape is not perfectly selective, as there is more than just a single undesired transition (see \cref{fig:DRAG}(a)). Despite this, this pulse shape can perform well in practice when $\Delta_1$ is treated as an optimization parameter. In fact, in \cref{fig:sel_pulse_breakdown} we see that these \texttt{exact 1-component DRAG} pulses with only the one optimization parameter achieve very similar performance to the \texttt{standard DRAG} pulses with three optimization parameters.

The \texttt{approximate 2-component DRAG} pulse exactly implements a given rotation at a desired transition, while approximately suppressing rotation at two undesired transitions with detunings $\Delta_{1},\Delta_2$. The pulse is defined as  
\begin{align}
\label{app_eq:approx_2_omega}
    \Omega(t) = \Omega_0(t)-i\frac{[E^2_1(t)-E^2_2(t)]\Delta_1\Delta_2}{E^2_1(t)E^2_2(t)(\Delta_1-\Delta_2)}\dot \Omega_0(t) \nonumber \\ -\frac{E^2_1(t)\Delta_2-E^2_2(t)\Delta_1}{E^2_1(t)E^2_2(t)(\Delta_1-\Delta_2)}\ddot\Omega_0(t),
\end{align}
In the above, for $\Omega_0(t)$ we take the $m=2$ truncated Gaussian pulse, $E^2_i(t)=\Omega_0(t)^2+\Delta_i^2$, and the time-dependent pulse detuning and rotation angle are again given by \cref{app_eq:e1_delta,app_eq:e1_theta}. Cf Eq.~(5.15) in \cite{motzoi2013improving}, where here we have simplified by neglecting terms $\propto\dot E_i(t)$, which is justified in the far detuned limit $\Delta_i\gg \Omega_0$. In the context of the \cRX{} gate, $\Delta_1,\Delta_2$ are treated as optimization parameters.  

Finally, the \texttt{semiclassical 2-component DRAG} pulses are obtained by taking the $\Omega_0(t)/\Delta_i\to 0$ limit of the \texttt{approximate 2-component DRAG} pulses, yielding
\begin{align}
    \Omega(t)= \Omega_0(t)-i\left(\frac{1}{\Delta_1}+\frac{1}{\Delta_2}\right)\dot\Omega_0(t)-\frac{1}{\Delta_1\Delta_2}\ddot\Omega_0(t)
\end{align}
with time-dependent pulse detuning and rotation angle given by \cref{app_eq:e1_delta,app_eq:e1_theta}.  In \cref{fig:sel_pulse_breakdown} we see that the optimized \texttt{approximate 2-component DRAG} and \texttt{semiclassical 2-component DRAG} pulses significantly outperform the optimized pulses \texttt{standard DRAG} and \texttt{exact 1-component DRAG}. The \cRX{} gate performance when using the \texttt{semiclassical 2-component DRAG} pulse is plotted as the orange ``standard DRAG'' line in \cref{fig:DRAG}.

Finally, for completeness' sake, we note that the 2-component pulses can be extended to more than two undesired transitions. The full expression for the 3-component approximate pulse is quite involved, but the corresponding 3-component semiclassical pulse takes the comparatively simple form
\begin{align}
    \Omega(t) &= \Omega_0(t)-i\left(\frac{1}{\Delta_1}+\frac{1}{\Delta_2}+\frac{1}{\Delta_3}\right)\dot\Omega_0(t)\nonumber\\
    &-\frac{\Delta_1+\Delta_2+\Delta_3}{\Delta_1\Delta_2\Delta_3}\ddot\Omega(t)+i\frac{1}{\Delta_1\Delta_2\Delta_3}\frac{d^3\Omega_0(t)}{dt^3},
\end{align}
where for $\Omega_0(t)$ we take the $m=3$ truncated Gaussian pulse.
However, in \cref{fig:sel_pulse_breakdown} we see that these 3-component pulses do not offer any further improvement over the relevant range of parameters.

\subsection{Additional applications of the \cRX{} gate}
\label{app:crx_applications}

Here we show that the \cRX{} gate naturally enables other useful capabilities, including high-fidelity, bias-preserving cat $Z$ and cat-cat $CZ$ rotations, as well as single-shot cat $Z$ basis readout.

Cat $Z$ rotations can be implemented by exploiting the phase accumulation associated with the \cRX{} gate's selective pulses. For example, when the transmon is initialized in $\ket{+}$, the \cRX{} unitary
\begin{equation}
    CR_X = \ket{\alpha}\bra{\alpha} \otimes I + \ket{-\alpha}\bra{-\alpha} \otimes R_X(\pi)
\end{equation}
acts trivially on the transmon. Since the transmon and cat are not entangled, the transmon can be discarded, and the net effect of the \cRX{} gate is to enact an $S = \text{diag}(0,i)$ gate on the cat. The $S$ gate is equivalent to a $Z$ rotation by $\pi/2$. In fact, it is possible to implement arbitrary cat $Z$ rotations in this manner. Tuning the selective pulses so that the transmon starts and ends in the same state while acquiring a phase $\phi$ (see, e.g., Refs.~\cite{krastanov2015universal,li2020approach}) is equivalent to enacting a cat $Z$ rotation by $\phi$. 

One can similarly implement cat-controlled cat-$Z$ rotations via a generalized \cRX{}. Consider two cats dispersively coupled to the same transmon. Displacing both cats and applying selective pulses enables transmon rotations conditioned both cats being in their logical $|1\rangle$ states. Tuning the selective pulses so that the transmon accumulates a $\phi$ phase is equivalent to the multi-cat rotation $\text{diag}(1,1,1,e^{i\phi})$, equivalent to a CZ gate for the choice of $\phi = \pi$. This approach easily generalizes to $n>2$ cats.  

What is remarkable about the ability of the \cRX{} to implement cat $Z$-type rotations is that this implementation is exponentially noise-biased (since the \cRX{} gate is itself exponentially noise biased), yet the infidelity is not subject to the unfavorable $\sqrt{\kappa_1/\kappa_2}$ scaling of prior bias-preserving approaches to $Z$ rotations based on the Zeno effect~\cite{mirrahimi2014dynamically,guillaud2019repetition}. High-fidelity and high-bias $Z$ rotations are thus possible with this approach. 
Ref~\cite{gautier2023designing} similarly proposes ways to circumvent the unfavorable infidelity scaling. In comparison to the proposals in that work, this \cRX{}-based approach is practically appealing in the context of $Z$ rotations, but it is not obviously generalizable to enable a high-fidelity cat-cat \cX{} gate.

Finally, \cRX{} gate can also be used to enable single-shot $Z$ basis readout of cat qubits with readout error that decreases exponentially with $|\alpha|^2$. Readout is performed by initializing the transmon in $\ket{g}$, applying the \cRX{} gate, then reading out the state of the transmon. An outcome of $\ket{g}$ indicates the cat logical $\ket{0} (=\ket{+\alpha})$, and an outcome of $\ket{e}$ indicates cat logical $\ket{1} (=\ket{-\alpha})$. A simplified version of this readout was employed in Ref.~\cite{repcat}. This readout procedure does not corrupt the cat's logical $Z$ information, and because of this non-demolition property the readout can be repeated to improve performance. (Note that the dissipative cat stabilization can be turned on during the transmon readout, i.e.~between \cRX{} gates, so that leakage out of the code space does not accumulate with increasing repetitions.) In particular, repeating the procedure and taking the majority vote allows one to exponentially suppress contributions to the overall cat readout error associated with transmon ancilla errors~\cite{hajr2024high}, at the cost of a longer effective readout duration. With sufficiently many repetitions, the overall readout error is thus only limited by the cat's exponentially small (in $|\alpha|^2$) intrinsic bit-flip probability. 

\section{Pulsed stabilization}
\label{app:pulsed_stabilization}
\begin{figure}
    \centering
    \includegraphics[width=0.8\linewidth]{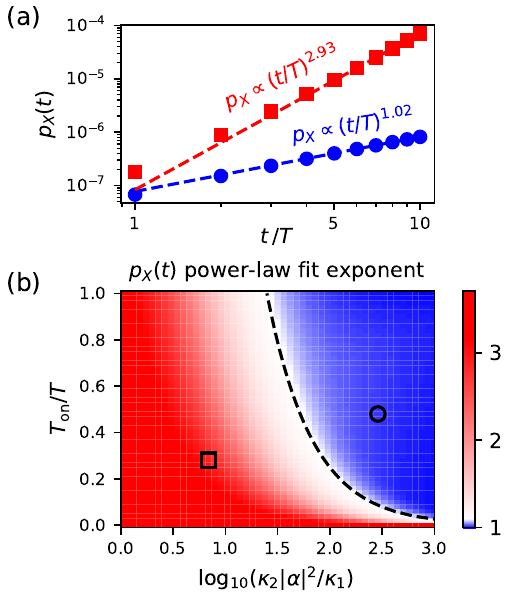}
    \caption{Efficacy of pulsed stabilization.
    (a) Cumulative bit-flip probability scaling. The cumulative probability of a bit flip, $p_X(t)$, is plotted (symbols) as a function of time and fit (dashed lines) to a power law. The different colors correspond to different choices of $T_\mathrm{on}/T$ and $\kappa_2|\alpha|^2/\kappa_1$. (b) Fitted power-law exponents. The data in (a) correspond to circular and square markers. The black dashed line is $\frac{\kappa_2 |\alpha|^2}{\kappa_1}\frac{T_\mathrm{on}}{T} \geq 25$; the good agreement of this curve with the contour of the plot indicates that the effect of pulsing can be well approximated via a reduced stabilization rate: $\kappa_2\to\kappa_2 (T_\mathrm{on}/T)$. In the blue region of parameter space, the bit-flip probability increases approximately linearly with time, indicating well-stabilized cats. }
    \label{fig:pulsed_stabilization}
\end{figure}

In this section we quantify when pulsed stabilization is sufficient to prevent accumulating leakage out of the code space so that cats retain well-defined bit-flip rates. \cref{fig:pulsed_stabilization} shows results from master equation simulations of a cat subject to single-photon loss, where stabilization is repeatedly pulsed on and off. The pulsed stabilization has duty cycle $T_\mathrm{on}/T$, where $T_\mathrm{on}$ is the time the stabilization is on, and $T = T_\mathrm{on} + T_\mathrm{off}$ is the duration of a single on-off pulsing round. The results show that if the stabilization is too weak or the duty cycle too small, the bit-flip probability grows nonlinearly in time, indicating accumulating leakage. In contrast, when both the stabilization rate and duty cycle are sufficiently high, then the cat exhibits a well-defined (i.e.~constant) bit-flip rate, indicating no leakage accumulation. To identify the regime of well-stabilized cats, we compare the single photon loss rate $\kappa_1$ against the effective confinement rate, $ \propto \kappa_2 |\alpha|^2 T_\text{on}/T$, where this effective confinement differs from the usual confinement rate, $\propto \kappa_2|\alpha|^2$~\cite{lescanne2020exponential}, by a factor of the duty cycle in order to account for the pulsing. We find that $\frac{\kappa_2 |\alpha|^2}{\kappa_1}\frac{T_\mathrm{on}}{T} \gtrsim 25$ is sufficient to guarantee a well-stabilized cat. 

For context, consider the example of $\chi/2\pi = 2\,\text{MHz}$, corresponding to \cX{} and \cRX{} gate durations of $\pi/\chi = 250\,\text{ns}$. Assuming the same duration for transmon readout and reset, the pulsed stabilization duty cycle is $T_\text{on}/T=1/5$ during surface code error correction (longer transmon readout and reset duration would only yield a larger duty cycle). Assuming loss rates in the vicinity the surface code ``threshold'' identified in \cref{sec:error_threshold}, $\kappa_1 = (100\mu s)^{-1}$, and $|\alpha|^2 = 4$, we find that $\kappa_2/2\pi \lesssim 50$kHz is required for a well-stabilized cat. This requirement has already been exceed by current experiments~\cite{reglade2024quantum,marquet2024autoparametric,singlecat}, and the requirement only becomes more lenient as $\kappa_1$ decreases of $|\alpha|^2$ increases. Stabilizing the cats only during the ancilla transmons' readout and reset is thus a realistic way to prevent accumulating leakage. That said, additional stabilization steps could always be inserted between cat-transmon entangling gates if needed, at the price of longer syndrome extraction rounds.

\section{Alternative noise models}
\label{app:alternative_noise_models}

In order to investigate the dependence of our results on the choice of noise model, in this section we consider variants of the noise model of \cref{tab:noise_model} of main text. The three noise models we consider are given in \cref{tab:many_noise_model}. Model 1 is the same as that in \cref{tab:noise_model} and is repeated here for ease of comparison. Recall that this model assumes higher coherence for the storage mode than the transmon, which is roughly representative of 3D devices, as well as the specific 2D device of Ref.~\cite{repcat}. In contrast, Model 2 assumes that the storage mode and transmon have comparable coherence (storage $T_1$ = transmon $T_1$), which may be a more reasonable assumption for 2D devices generally. Separately, in order to provide insights into the limitations imposed by transmon decoherence, in Model 3 the transmon decoherence rates are held fixed, unlike in Models 1 and 2 where they are varied together with the storage deocherence rates.

\begin{table}[]
    \centering
    \begin{tabular}{|l|l|l|l|}
    \hline
        Decoherence source & Model 1 & Model 2 & Model 3 \\
        \hline
        \hline
        Storage loss $(\kappa_1/\chi)$ & $q$ & $q$ & $q$ \\
        \hline
        Transmon loss $(\gamma_\downarrow/\chi)$ & $3q $ & $q$ & $5\times 10^{-4}$ \\
        \hline
        Storage dephasing $(\kappa_\phi/\chi)$ & $0.01 q$ & $0.01q$ & $0.01 q$ \\
        \hline
        Transmon dephasing $(\gamma_\phi/\chi)$ & $1.5q$ & $0.5q$ & $2.5\times 10^{-4}$\\
        \hline
        Transmon heating $(\gamma_\uparrow/\chi)$ & $0.015q$ & $0.005q$ & $2.5\times 10^{-6}$ \\
        \hline
        
    \end{tabular}
    \caption{Different single-parameter noise models used to obtain the results in \cref{fig:achievable_bias_pZ_many_models}. Model 1 is the same as in \cref{tab:noise_model} of the main text. Relative to this, Model 2 reduces all transmon decoherence rates by a factor of 3, and Model 3 holds the transmon decoherence rates constant (i.e.~independent of $q$). The ratios between transmon loss and dephasing and heating rates are maintained, so all models assume an effective thermal population of $\gamma_\uparrow/\gamma_\downarrow = 0.5\%$ and $T_1 = T_2$ for the transmon.}
    \label{tab:many_noise_model}
\end{table}

\begin{figure*}
    \centering
    \includegraphics[width=1\linewidth]{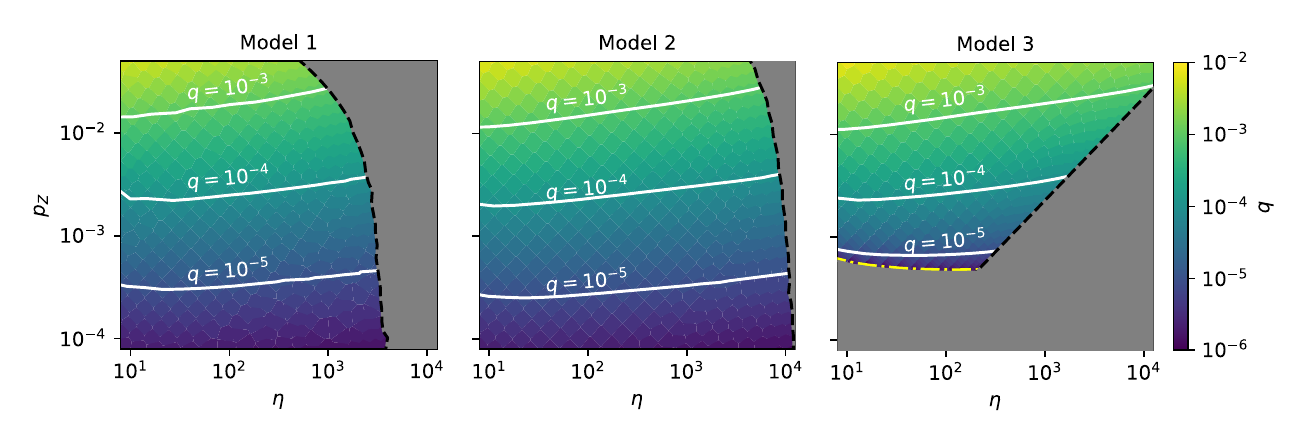}
    \caption{Achievable noise bias, $\eta$, and $Z$-type error probability, $p_Z$, for the different noise models in \cref{tab:many_noise_model}. Gray areas indicate regions of parameter space that are either inaccessible or where operation is not beneficial to logical memory performance. Black dashed lines indicate the maximum beneficial bias (see \cref{App:optimal_alpha}). The yellow dot-dashed line in the Model 3 plot indicates the limit imposed on $p_Z$ by the fixed transmon decoherence. This line is obtained by computing $p_Z$ in the limit of zero storage decoherence (i.e.~$q\to 0$ in Model 3). }
    \label{fig:achievable_bias_pZ_many_models}
\end{figure*}

In \cref{fig:achievable_bias_pZ_many_models}, we plot the achievable noise bias ($\eta$) and $Z$-type error probability ($p_Z$) for these three different noise models,  as in \cref{fig:achievable_bias_pZ} of the main text. 
Starting with Model 1, the corresponding plot in \cref{fig:achievable_bias_pZ_many_models} is the same as \cref{fig:achievable_bias_pZ}, except that we have grayed out the region of parameter space that lies beyond the maximum beneficial bias. This is done both to emphasize that operating in this regime of parameter space, while technically possible, only serves to degrade logical performance (see \cref{App:optimal_alpha}), and also to visually highlight the contrast with the other noise models. Indeed, for Model 2, the main difference relative to Model 1 is that larger biases are accessible. The increase in the accessible bias follows simply from the lower bound on the cat bit-flip probability of the \cX{} gate (\cref{eq:transmon_bit_flip_limit}). Since transmon errors during the \cX{} gate can induce cat bit flips, higher transmon coherence relative to storage mode coherence allows for a larger accessible bias. For example, with transmon and storage $T_1$ both on the scale of milliseconds and $\chi$ of the scale of MHz (corresponding to $q$ in the range $10^{-5}-10^{-4}$), biases of $10^{4}$ are achievable. This is a roughly a factor of $3$ larger than the maximum achievable bias for Model 1 at the same $q$, consistent with the fact that the transmon heating rate is $3\times$ lower in Model 2. 

Model 3 holds the transmon decoherence rates fixed, so the corresponding plot in \cref{fig:achievable_bias_pZ_many_models} highlights the limitations on $p_Z$ and $\eta$ imposed by transmon decoherence. Specifically, the fixed values of $\gamma_\downarrow/\chi=5\times 10^{-4}$ and $\gamma_\phi/\chi = 2.5\times 10^{-4}$ correspond to per-gate transmon decay and dephasing probabilities on the order of $10^{-3}$. Because transmon errors can propagate through the \cRX{} gate to cat $Z$ errors, the achievable $p_Z$ is limited to a similar value. This limitation is indicated by the dot-dashed yellow line in the plot, making clear that after a point improving storage coherence (decreasing $q$) yields no improvement in $p_Z$, unless transmon coherence is also improved. 
At the same time, with fixed transmon decoherence, the maximum beneficial bias scales linearly with $p_Z$ (black dashed line). This scaling results from the fact that the fixed transmon decoherence fixes the minimum achievable cat bit-flip probability of the \cX{} gate (\cref{eq:transmon_bit_flip_limit}), and with fixed bit-flip probability we have $\eta \propto p_Z$.
Both this scaling, and a visual comparison of the achievable bias with Models 1 and 2 to that with Model 3, make clear that it is necessary to proportionately increase the transmon coherence in order to maintain a large noise bias as the storage mode coherence is increased.

\section{Extracting gate error channels}
\label{app:extracting_error_channels}

We numerically extract Pauli error channels describing \cX{} and \cRX{} gates as follows. The gate error $\mathcal{E}$ channel is defined such that the application of a noisy gate is equivalent to the ideal gate followed by application the error channel $\mathcal{E}$,
\begin{equation}
    \mathcal{U}_\text{noisy}(\rho) = \mathcal{E}(U_\text{ideal}\rho U_\text{ideal}^\dagger).
\end{equation}
We can characterize the error channel $\mathcal{E}$ via its Pauli transfer matrix, $R$, with matrix elements given by
\begin{equation}
    R_{ij} = \frac{1}{d}\text{tr}[P_i \mathcal{E}(P_j)],
\end{equation}
where $d=4$ is the Hilbert space dimension of two qubits, and the $P_{i}$ denote (two-qubit) Pauli operators. This representation is convenient to extract from simulations, as one calculate $R_{ij}$ by initializing the system in $U^\dagger_\text{ideal} P_j U_\text{ideal}$, applying the noisy gate operation $\mathcal{U}_\text{noisy}$, and computing the overlap with $P_i$,
\begin{equation}
    R_{ij}  = \frac{1}{d} \text{tr}[P_i \mathcal{U}_\text{noisy}(U^\dagger_\text{ideal} P_j U_\text{ideal})].
\end{equation}

For error correction simulations, a more convenient representation is the chi-matrix (or process matrix) representation,
\begin{equation}
    \mathcal{E}(\rho) = \sum_{i,j} \chi_{i,j} P_i \rho P_j^\dagger.
\end{equation}
To enable efficient error correction simulations, we make the Pauli twirling approximation and neglect the off-diagonal terms in $\chi$. The remaining diagonal terms encode the probabilities of different Pauli errors, and they are related to the Pauli transfer matrix via a Walsh Hadamard transform,
\begin{equation}
    R_{ii} = \frac{1}{d} \sum_{j,k} \chi_{j,k} \text{tr}[P_i P_j P_i P_k] = \sum_j \chi_{jj}(-1)^{\left<i, j\right>},
\end{equation}
where $\left<i, j\right>$ is defined via $P_i P_j = (-1)^{\left<i, j\right>} P_i P_j$. We can thus compute the diagonal chi-matrix elements (Pauli error probabilities) by applying an inverse Walsh Hadamard transform to the diagonal Pauli transfer matrix elements obtained from gate simulations.

In noisy implementations of the cat-transmon gates, errors can prevent the system from returning to the code space. That is, the storage mode state could lie outside the cat code space (spanned by $\ket{0/1}_C$), and, in the case of the \cX{} gate, the final transmon state could lie outside the gate's $\ket{g},\ket{f}$ code space. This possible leakage must be accounted for to obtain a normalized Pauli error channel for the gate, and we do so by augmenting the noisy gate channel with a recovery operation: $\mathcal{U}_\text{noisy} \to \mathcal{R} \circ \mathcal{U}_\text{noisy}$, where the recovery operation $\mathcal R$ operation returns the system to the code space as described below.

For the storage mode, $\mathcal R$ implements the ideal dissipative recovery map, which is equivalent to applying the dissipative stabilization $\mathcal{D}[a^2-\alpha^2]$ for infinite time. We implement this map by computing the conserved quantities described in \cite{mirrahimi2014dynamically}. This choice of recovery operation is justified by the periodic application of the dissipative stabilization during error correction cycles (see \cref{fig:architecture}). which suffices to prevent the accumulation of leakage outside the code space as shown in \cref{app:pulsed_stabilization}. 

For the transmon, $\mathcal R$ acts differently depending on whether the transmon is operated in its $\ket{g},\ket{e}$ subspace (\cRX{} gate) or its $\ket{g},\ket{f}$ subspace (\cX{} gate). In the case of the \cRX{} gate, $\mathcal R$ acts trivially as there is no leakage outside the code space. In the case of the \cX{} gate, $\mathcal R$ returns any leaked population in $\ket{e}$ to $\ket{f}$. This is the only choice for $\mathcal{R}$ that accurately captures the resilience of the \cX{} gate to single transmon $T_1$ decay events; were any $\ket{e}$ population mapped to $\ket{g}$, single decay events ($\ket{f}\to\ket{e}$) would effectively be mapped to double-decay events ($\ket{f} \to \ket{e} \to \ket{g}$), and the moderately noise-biased nature of the gate would not manifest in the Pauli error channel.

It is worth highlighting that the this approach---computing gates' Pauli error channels by applying a recovery $\mathcal{R}$ after the simulation of a single noisy gate---does not capture the possible impacts of leakage propagating between subsequent gates. While it is true that the cats' dissipative stabilization prevents the accumulation of leakage, multiple gates are applied in the stretches between dissipative stabilization (see \cref{fig:architecture}), so leakage in an earlier gate can in principle impact the performance of a later gate before it is corrected by the stabilization. For the cats' themselves, neglecting this effect is unlikely to have significant impact. In particular, small distortions in the input cat states should not propagate through \cX{} or \cRX{} gates to large enough distortions to induce cat $X$ errors. Indeed, no such damaging propagation was observed in the experiment~\cite{repcat} (see also the fault-tolerance arguments in Ref.~\cite{xu2024fault}). Were propagation of leakage an issue, the dissipative stabilization could be applied more frequently~\cite{regent2022high}, at the price of a longer error correction cycle duration.  

Separately, for the transmons, neglecting the propagation of leakage does omit the possibility that an $\ket{f}\to\ket{e}$ decay in one \cX{} gate could combine with an $\ket{e}\to\ket{g}$ decay in a later \cX{} to induce a cat $X$ error. Treating leakage as such thus modestly underestimates the probability of double-decay-induced cat $X$ errors. This underestimate will not affect results qualitatively, as double-decay-induced cat $X$ errors are accounted for within individual gates. It is also unlikely to have a significant quantitative impact, since for sufficiently high coherence the contribution from transmon double decay (scales quadratically in the noise parameter $q$), will be negligible in comparison to that from transmon heating (scales linearly in $q$), and the impacts of heating are accurately captured by this approach.  

\section{Simulation assumptions}
\label{App:sim_assumptions}

In this section, we enumerate various assumptions that are made implicitly in our simulations of the cat-transmon architecture. For each, we discuss when the assumption is justified and its potential impacts on performance. 

\textbf{Perfect $\chi$ matching.} Our simulations of the \cX{} gate assume $\chi_e = \chi_f$. In practice, small mismatches $|\chi_e/ \chi_f - 1| \neq 0$ can be tolerated without meaningfully increasing the cat $X$ error rate. This tolerance results from the periodic application of engineered dissipation, which is capable of correcting the small storage mode over- or under-rotations associated with the mismatch, as demonstrated in Ref.~\cite{repcat}. What constitutes a tolerable mismatch generally depends on the engineered dissipation strength and how frequently it is applied. For context, Ref.~\cite{repcat} found mismatches as large as $\approx 20\%$ did not meaningfully increase the cat $X$ error rate with stabilization applied only once per error correction cycle. So long as the mismatch is not too large to significantly degrade the performance of a single \cX{} gate, engineered dissipation can in principle be applied more frequently (e.g.~after each \cX{} gate) to ensure that the mismatch does not result in a significantly increased cat $X$ error rate. This added protection, however, would come at the price of a longer error correction cycle. 

\textbf{No tunable coupler heating.} In implementations of the cat-transmon architecture that rely on tunable couplers to implement the tunable cat-transmon dispersive coupling~\cite{repcat}, heating of the couplers can potentially induce cat $X$ errors just as can heating of the transmon errors. This coupler-heating-induced cat $X$ error probability is negligible if the coupler's dispersive coupling to the storage modes is sufficiently small in comparison to the transmon's. However, if the coupler's dispersive coupling is comparable or larger than the transmon's, coupler heating can contribute appreciably to the cat $X$ error probability. The impact of this additional contribution can be effectively assessed from our results simply by augmenting the transmon heating rate, indicating that coupler heating should not qualitatively impact results.

\textbf{No storage mode self Kerr.} Typically, dispersively coupling a storage mode to a nonlinear transmon qubit causes the storage mode to acquire a self-Kerr nonlinearity. The magnitude of this self Kerr will depend on how the coupling is engineered (e.g.~the proposal of Ref.~\cite{ye2021engineering} realizes a dispersive coupling without any storage self Kerr), but small self Kerr can be tolerated by the cat-transmon architecture without meaningfully degrading performance. This tolerance again stems predominantly from the periodic application of engineered dissipation, which is capable of correcting small distortions to the cats caused by self Kerr. For example, Refs.~\cite{singlecat, repcat} found that self Kerrs on the order of a few kHz were tolerable. We note also that the \cRX{} gate's storage echo effectively mitigates the prospect of cat $Z$ errors induced by the storage self Kerr.

\textbf{Two-level transmon approximation for the \cRX{} gate}. We only include the lowest two transmon levels in our simulations of the \cRX{} gate. This approximation is valid, i.e.~leakage to higher states like $\ket{f}$ is likely to be negligible, when the \cRX{} gate's pulses are sufficiently selective. In particular, if the bandwidth of the selective pulse is much smaller than the transmon anharmonicity, the selective pulse should not induce leakage. As a rule of thumb, the bandwidth of the selective pulses must be smaller than $4|\alpha|^2 \chi$ for good \cRX{} performance, so leakage to higher states should be negligible when the anharmonicity is larger than this quantity. For MHz-scale $\chi$, this assumption is reasonable, as typical transmon anharmonicities are on the order of 200 MHz. The assumption may be violated, however, if larger $\chi$ is used in an effort to speed up gates. In this situation, DRAG can be used to mitigate transmon leakage during the \cRX{} gate's selective pulses.

\textbf{Perfect coupler toggling.} Our simulations assume that the tunable dispersive couplings can be instantly toggled from on to off, with zero dispersive coupling at the off position, whereas in actuality toggling the coupling will require some finite duration and there will be a finite on-off ratio. The primary impact of the finite switching time is to effectively lengthen the duration of the cat-transmon gates. This is especially relevant for the \cRX{} gate, which ideally involves toggling the coupler on and off four times during the gate sequence. This additional time cost could be offset by operating the \cRX{} gate at larger $|\alpha|^2$, allowing for faster selective pulses at the price of increased cat $Z$ error. Alternatively, one could leave the coupler on for the entire gate, relying on wide-bandwidth pulses to implement the unselective transmon pulses and storage displacements accurately even in the presence of non-negligible dispersive coupling. The primary impact of the finite on-off ratio is to induce small distortions of the cat, which are again correctable via periodic application of the engineered dissipation, as demonstrated experimentally in Refs.~\cite{singlecat, repcat}.

\section{Fitting and extrapolating logical errors}
\label{App:logical_err_fitting}
\begin{figure*}
    \centering
    \includegraphics[width=\textwidth]{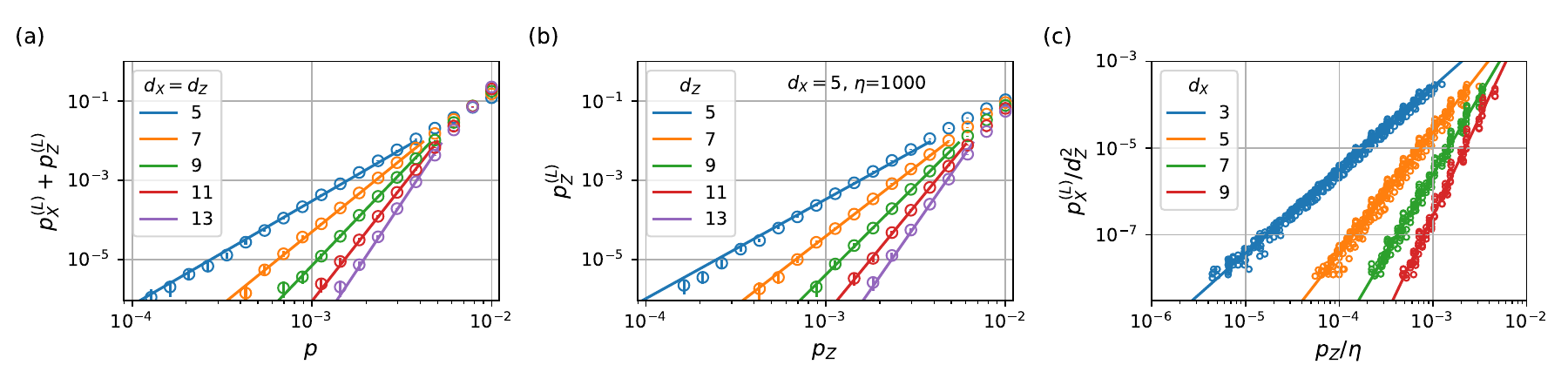}
    \caption{Logical error projections for biased and unbiased noise models. \textbf{(a)} Total logical error and fits for square ($d_X=d_Z\equiv d$) surface codes for the unbiased error model.  \textbf{(b)} Logical Z error for the biased error model, for fixed $d_X$ and $\eta$. \textbf{(c)} Logical $X$ error for the biased error model.  Error bars in (a) and (b) indicate 95\% confidence intervals; error bars are omitted in (c) for visual clarity. }
    \label{fig:logical_fitting}
\end{figure*}

Our procedure for fitting and extrapolating logical errors is illustrated in \cref{fig:logical_fitting}. For unbiased noise [\cref{fig:logical_fitting}(a)], we fit the total logical error  data (aggregated across all $d$) to the ansatz $p_X^{(L)} + p_Z^{(L)} = Ad^2(Bp)^{Cd}$, where $d_Z = d_X = d$, and capital letters are fit parameters, with the resulting fit shown as solid lines. The $d^2$ prefactor accounts for the fact that the number of fault locations that contribute to logical $X$ ($Z$) errors increases both with the number of rounds ($d$), and the distance $d_Z$ $(d_X)$. This is only an approximation (see more detailed discussion in, e.g., Ref.~\cite{fowler2012surface}), but the ansatz nevertheless fits the data well.  

For biased noise, we separately fit $p_Z^{(L)}$ and $p_X^{(L)}$. The $p_Z^{(L)}$ data [\cref{fig:logical_fitting}(b)] is fit to the ansatz $A d_Z (B p_Z)^{C d_Z}$, with fitting performed separately for each different choice of $d_X$, $\eta$. The prefactor of $d_Z$ accounts for the fact that we simulate $d_Z$ rounds of error correction, and we opt not to include an additional factor of $d_X$ in the prefactor, since $d_X$ is held fixed for each fitting. To fit $p_X^{(L)}$ [\cref{fig:logical_fitting}(c)] , data is generated from logical memory simulations with a variety of choices for $p_Z$, $\eta$, and $d_Z$. For fixed $d_X$, we find that $p_X^{(L)}/d_Z^2$ is well described by a power law fit of the form $A(p_Z/\eta)^B$. The $d_Z^2$ accounts for the increasing number of fault locations, and $p_Z/\eta$ is a proxy for the physical $X$ error probability.

\begin{figure}
    \centering
    \includegraphics[width=1.0\linewidth]{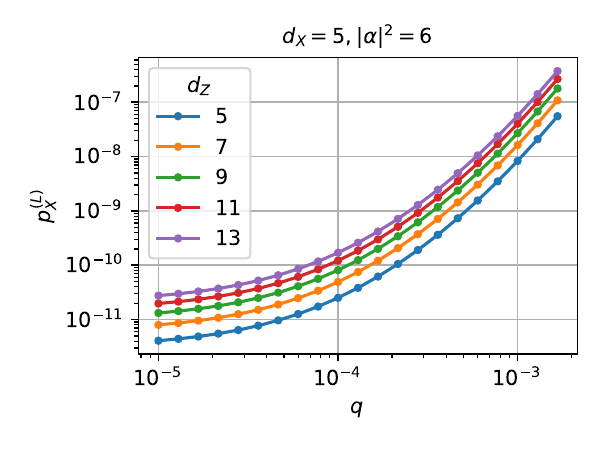}
    \caption{Extrapolated $p_X^{(L)}$ for fixed $d_X=5$ and $|\alpha|^2 = 6$. Gate simulations are used to compute $p_Z$ and $\eta$ over a range $q$ values (using the noise model in \cref{tab:noise_model}), then $p_X^{(L)}$ is computed using the $d_X=5$ fit from \cref{fig:logical_fitting}(c). }
    \label{fig:pXL_dX5}
\end{figure}

In \cref{fig:pXL_dX5}, we use the fits of $p_X^{(L)}/d_Z^2$ to verify the claim made in the caption of \cref{fig:threshold} that $d_X = 5$ and $|\alpha|^2 = 6$ is sufficient to achieve $p_X^{(L)} \lesssim 10^{-10}$ in the deep sub-threshold regime.  Specifically, we simulate the \cX{} and \cRX{} gates at fixed $|\alpha|^2 = 6$ and varied $q$, then extract $p_Z$ and $\eta$ from the gate error channels. We then use the fit of $p_X^{(L)}/d_Z^2$ to estimate $p_X^{(L)}$ at various $d_Z$. Given the ``threshold'' of $q=5\times10^{-4}$ in \cref{fig:threshold}, in the deep sub-threshold regime ($q\leq 5\times 10^{-5}$) we find that indeed $p_X^{(L)}< 10^{-10}$ for the $d_Z$ values shown. 

\section{Validating the simplified biased error model}

In this section, we compare the logical memory error of rectangular surface codes under two different circuit-level noise models: the ``full model,'' where Pauli error channels are directly extracted from master equation simulations of the cat-transmon gates, and the ``simplified model'' that replaces these error channels with simply parameterized biased error channels. 

The two parameters describing the simplified error channels are the $Z$ error probability, $p_Z$, and the bias $\eta$. We compute these parameters from the full \cX{} and \cRX{} gate Pauli error channels as follows. We define
\begin{equation}
    p_Z = \frac{3}{2}\times\frac{1}{2}\left[p_{ZI}^{cX}+p_{ZZ}^{cX}+p_{ZI}^{cRX}+p_{ZZ}^{cRX}\right]
\end{equation}
where the factor of $3/2$ is a phenomenological factor that we find is necessary to obtain good agreement, and the remainder of the expression is the average per-gate probability of a $Z$ error on a given cat qubit, averaged over the \cX{} and \cRX{} gates. We then define the bias as
\begin{equation}
    \eta = p_Z/(p_X+p_Y)
\end{equation}
Here, $(p_X+p_Y)$ denotes the sum of all probabilities for $X$ or $Y$ errors on the cat qubit, averaged over the \cX{} and \cRX{} gates. 
The simplified error error channels are then defined (for both single- and two-qubit gates) such that all $Z$-type errors have equal probability, with the total probability of a $Z$-type error equal to $p_Z$, and the total probability of any other error
equal to $p_Z/\eta$.

\label{App:validating_simplified_model}
\begin{figure*}
    \centering
    \includegraphics[width=\textwidth]{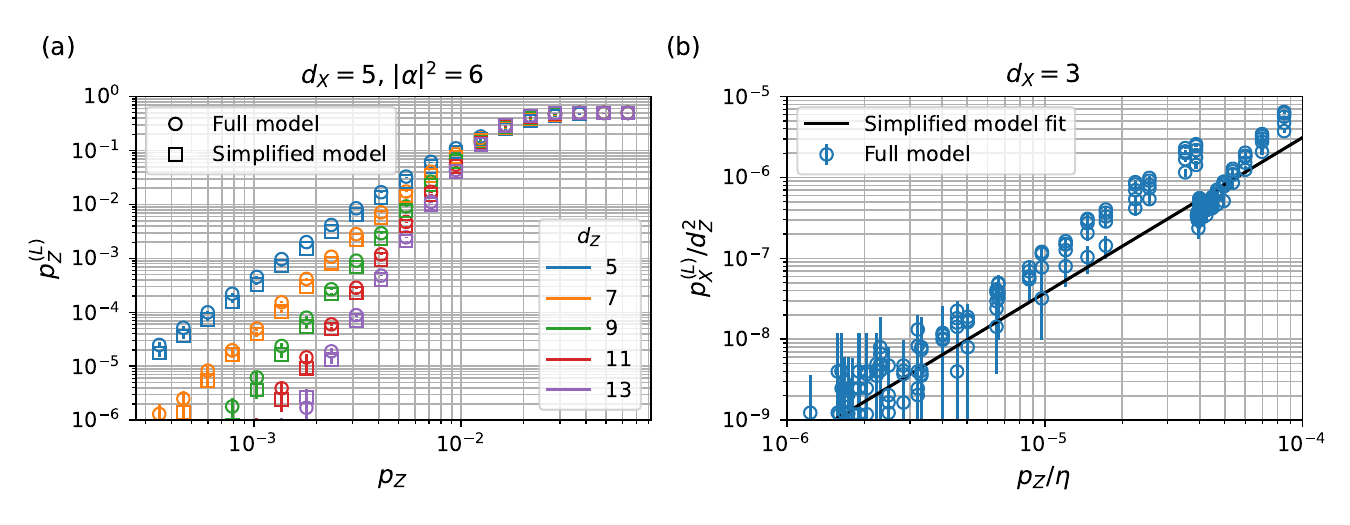}
    \caption{Comparing logical error rates for the full and simplified error models \textbf{(a)} Comparison of $p_Z^{(L)}$. \textbf{(b)} Comparison of $p_X^{(L)}$. $p_X^{(L)}$ is computed either directly using the full error model (points), or from extrapolating a fit (solid line, see  \cref{App:logical_err_fitting}) to the simplified error model data. The full error model data points comprise various different choices of the underlying parameters $\alpha$, $\kappa_1/\chi$, and $d_Z$, and the corresponding $p_Z,\eta$ are computed from the resultant \cX{} and \cRX{} error channels. All error bars indicate 95\% confidence intervals.  }
    \label{fig:simplified_model_verification}
\end{figure*}

With this mapping, we compare the logical memory performance with the two different circuit-level noise models in 
In \cref{fig:simplified_model_verification}.   We find that the logical error obtained with the simplified model is reasonably predictive of that obtained with the full model. Specifically, in \cref{fig:simplified_model_verification}(a), we compare the logical $Z$ errors, $p_Z^{(L)}$. We find good agreement between the  $p_Z^{(L)}$, with the agreement shown for the specific case of $d_X=5,|\alpha|^2=6$ being representative of the agreement for other choices of these parameters. In \cref{fig:simplified_model_verification}(b), we compare the logical $X$ errors, $p_X^{(L)}$. For all points, we find reasonable, order-of-magnitude agreement between the two, though for some points the simplified model fit falls outside the full model data error bars. The impact of a possible discrepancy in $p_X^{(L)}$ can be approximated as an effective rescaling of the noise bias. In particular, we expect $p_X^{(L)} \propto (1/\eta)^{(d_X+1)/2}$, so that an underestimate of $p_X^{(L)}$ by a factor $x$ is roughly equivalent to an overestimate of the bias by a factor $x^{-2/(d_X+1)}$. For example, at $d_X = 5$, an underestimate of $p_X^{(L)}$ by a factor of 2 ($x=1/2$) would be roughly equivalent to to an overestimate in the bias by a factor of $\approx 1.25$. \cref{fig:logical_overheads} provides a sense of how different levels of noise bias affect the logical memory overhead.

\section{Compensating the \cRX{} phase during error correction}
\label{App:cRX_phase}

\begin{figure}
    \centering
    \includegraphics[width=1\linewidth]{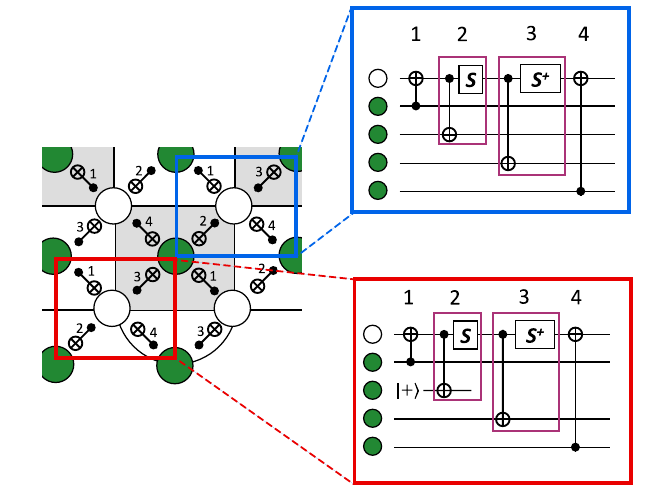}
    \caption{Cancellation of the \cRX{} gate's accompanying phase. The left panel shows a surface code patch with data cat qubits as white circles and ancilla transmon qubits as green circles. The \cX{} (green control, white target) and \cRX{} (white target) gates used for syndrome extraction are shown, with the time step during the error correction when each gate is applied indicated by the number above the gate. In the blue and red boxes on the right, the \cRX{} gates are represented by purple rectangles containing a standard \cX{} followed by an $S$ or $S^\dagger$ acting on the data cat qubit. These phase gates in adjacent time steps commute through the \cX{} gates and cancel as described in the text.  }
    \label{fig:cRX_S_compensation}
\end{figure}

As mentioned in the main text, the \cRX{} unitary
\begin{equation}
    CR_X = \ket{\alpha}\bra{\alpha} \otimes I + \ket{-\alpha}\bra{-\alpha} \otimes R_X(\pi),
\end{equation}
is equivalent to a cat-controlled \cX{} gate followed by an  $S^\dagger = \mathrm{diag}(1,-i)$ gate is to the cat. In \cref{fig:cRX_S_compensation}, we show that this additional phase gate does not present a problem for error correction. The basic idea is as follows. By changing the sign of the selective pulses in the \cRX{} pulse sequence, one can effectively replace $R_X(\pi)\to R_X(-\pi)$ in the above unitary, which is then equivalent to a \cX{} gate followed by $S$. We can then arrange for one \cRX{} gate's $S$ to be cancelled by another \cRX{} gate's $S^\dagger$ in the next time step of an error correction cycle, since the phase gates on the control commute with \cX{}. The net result is that we can effectively replace time-adjacent pairs of \cRX{} gates with pairs of \cX{} gates, and we recover the usual surface code syndrome extraction circuits. Importantly, this approach does not increase the length of an error correction cycle.

Special care needs to be taken when applying this approach to data qubits at the boundary, since in this case there is not guaranteed to be a \cRX{} at an adjacent time step. In this case (red box in \cref{fig:cRX_S_compensation}), an additional ancillary transmon qubit can be added outside of the surface code patch, and this qubit can be used together with a \cRX{} gate to apply an $S$ or $S^\dagger$ gate to the data cat qubit (see \cref{app:crx_applications}). In the context of fault tolerant computation with lattice surgery, such additional ancillary qubits would already be present in the associated routing space, so this strategy would not incur additional overhead.

\section{Comparison of CSS and XZZX surface codes}
\label{App:code_comparison}

\begin{figure*}
    \centering
    \includegraphics[width=1\linewidth]{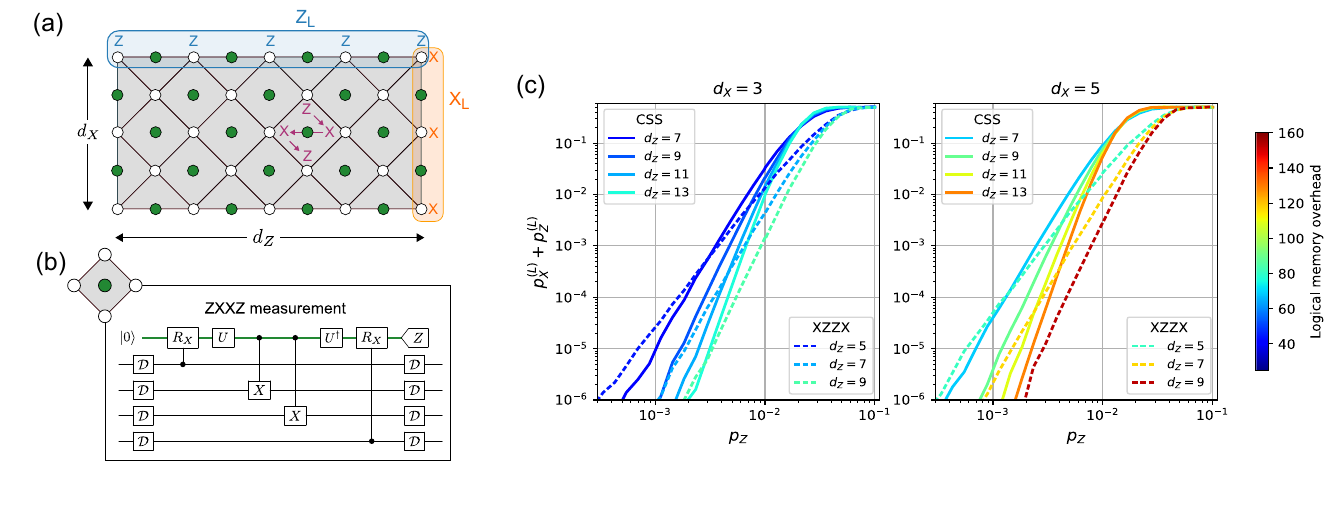}
    \caption{The XZZX code in the cat-transmon architecture. (a) Rectangular XZZX code. A $d_X = 3$, $d_Z = 5$  XZZX code is shown, comprising $(2d_X - 1)(2d_Z-1) = 45$ total qubits. The logical operators are shown in blue and orange. An example stabilizer generator is shown in purple, with arrows indicating the order in which entangling gates are applied between the data qubits (white circles) and ancilla qubit (green circle). (b) Syndrome extraction circuit. \cRX{} and \cX{} gates are used to enable measurement of $Z$-type and $X$-type Pauli operators, respectively. In between \cRX{} and \cX{} gates, a transmon operation $U$ is applied to convert from one transmon basis to another (see main text). (c) Comparison of rectangular XZZX (dashed lines) and CSS (solid lines) codes. We consider the simplified biased noise model described in \cref{Sec:logical} of the main text, with bias $\eta=10^3 $. Under this model, we plot the total logical error $p^{(L)} = p^{(L)}_X + p^{(L)}_Z$ after $d_Z$ rounds of noisy syndrome extraction for various code distances $d_X, d_Z$. For visual clarity, we plot only $d_X=3$ ($d_X = 5$) codes in the left (right) plots. Lines are colored according to their respective logical memory overheads: $2d_X d_Z - 1$ for the CSS code, and $(2d_X - 1)(2d_Z-1)$ for the XZZX code. For visual clairty, we omit plot markers and error bars: we simulate 30 uniformly space $p_Z$ values over the range $p_Z \in [10^{-4},10^{-1}]$, and we simulate repeated shots until 1000 logical errors are observed, up to $10^7$ total shots.}
    \label{fig:XZZX}
\end{figure*}

In this section, we consider a variant of the cat-transmon architecture where error correction is performed with the XZZX surface code~\cite{bonilla2021xzzx}, instead of with the standard ``CSS'' surface code described in the main text. As discussed in~\cite{bonilla2021xzzx} the XZZX code exhibits a higher threshold against highly biased noise in comparison to the CSS surface code. While the higher threshold allows for reduced logical memory overheads near threshold, we show that the CSS code nevertheless offers lower overheads in the practically-relevant deep sub-threshold regime, hence why we focus on the CSS code in the main text. 

The rectangular version of the XZZX code~\cite{bonilla2021xzzx,darmawan2021practical} is shown in \cref{fig:XZZX}(a). In contrast to the square version of the XZZX code, this rectangular version allows one to independently tune $d_X$ and $d_Z$ by adjusting the respective edge lengths of the code. This tunability allows for better hardware efficiency at large bias. For example, in the limit of inifinite bias $d_X = 1$ suffices, and the rectangular XZZX code reduces to a simple repetition code at $d_X = 1$. However, this rectangular XZZX code requires an ``unrotated'' lattice geometry, in contrast to the ``rotated'' lattice geometry of the rectangular CSS surface code in \cref{fig:architecture}(a). (The rotated lattice geometry is not suitable for rectangular XZZX codes: both $d_X$ and $d_Z$ would be limited by the length of the shorter side, due to the diagonal nature of the logical operators in the rotated geometry.) 
Owing to this rotation difference, the rectangular XZZX code requires more qubits to achieve the same code distance as the rectangular CSS code. In particular, the rectangular XZZX code comprises $(2d_X - 1)(2d_Z-1)$ total qubits (data and ancilla), which is roughly a factor of two larger than the overhead of the rectangular CSS code, $2d_X d_Z -1$. Thus, in comparison to the rectangular CSS code, the higher threshold of the XZZX code comes at the price of a larger overhead at fixed code distance. We analyze this tradeoff between higher threshold and larger overhead at fixed code distance below.   

In the cat-transmon architecture, XZZX error syndromes can be measured using \cX{} and \cRX{} gates with the circuit shown in \cref{fig:XZZX}(b). In this circuit, unitaries $U, U^\dagger$ are applied to the transmon between applications of the \cX{} and \cRX{} gates, where
\begin{equation}
    U = \pi_{e,f}H_{g,e}
\end{equation}
with $H_{g,e}$ denoting the Hadamard gate in the transmon's $\ket{g},\ket{e}$ subspace, and $\pi_{e,f} =\ket{f}\bra{e} + \text{H.c.}$ denoting an $\ket{e}\leftrightarrow\ket{f}$ $\pi$ pulse. This unitary serves two purposes: (1) it switches the transmon to its the complementary basis (via the Hadamard) so that it is sensitive to the appropriate type of error, and (2) it maps the transmon from its $\ket{g},\ket{e}$ subspace, where the \cRX{} gate is operated, to its $\ket{g},\ket{f}$ subspace, where the \cX{} is operated. Importantly, prior use of the $\ket{g},\ket{e}$ subspace does not undermine the noise bias of subsequent \cX{} gates. In particular, transmon errors before the first \cRX{} (caused, e.g., by a single decay event from $\ket{e}\to\ket{g}$) can propagate to an IXXZ error on the four connected data qubits. Because this error is stabilizer equivalent to ZIII, in effect this transmon error does not propagate to cat $X$ errors. Additionally, we note that the additional phase associated with the \cRX{} unitary (not shown in \cref{fig:XZZX}) can be compensated as described in \cref{App:cRX_phase} by cancelling the phase of a \cRX{} at the end of one round with that of a \cRX{} at the beginning of the next round.

In \cref{fig:XZZX}(c), we numerically compare the error correction performance of the CSS and XZZX surface codes. We use the ``simplified'' biased noise error model of \Cref{Sec:logical}, with $p_Z$ varied and the bias fixed to $\eta = 10^3$. This choice of $\eta$ is motivated by the fact that (i) it is achievable with the cat-transmon architecture, (ii) though the thresholds of both codes improve with bias, the thresholds saturate once $\eta\gtrsim 10^{3}$~\cite{bonilla2021xzzx}.
As expected, the XZZX code exhibits higher thresholds under this biased noise model. 
As can be seen by comparing XZZX and CSS curves of similar color in \cref{fig:XZZX}, the XZZX code offers lower logical error at fixed qubit overhead for $p_Z$ values near the threshold as a result. 
However, again comparing similar overheads (similar colors), we observe steeper slopes for the CSS code. 
As a result, for $p_Z$ far below threshold the CSS code exhibits lower logical error at fixed qubit overhead.  
Since in this work we are primarily interested in the deep sub-threshold regime, we thus focus on the CSS code. 

To provide intuition for this result, we note that the choice between the XZZX and CSS codes at fixed overhead can also be understood as a choice between increased threshold or increased code distance. That is, the XZZX code offers a higher threshold, but the CSS code offers a higher distance at fixed qubit overhead. At sufficiently low error, a higher code distance is always preferable to a higher threshold, as can be seen from the following scaling argument. Roughly speaking, logical error scales as $\sim (p/p_\text{th})^{d}$. Increasing the threshold $p_\text{th}\to p_\text{th}'>p_\text{th}$ suppresses logical error by a factor of $(p_\text{th}/p_\text{th}')^d$, while increasing the distance $d\to d'>d$ suppresses logical error by a factor of $(p/p_\text{th})^{d'-d}$. The latter is $p$-dependent, while the former is not. Thus, at some sufficiently small $p$, increasing the distance will always yield lower logical error than increasing the threshold, hence why the CSS code outperforms the XZZX code at sufficiently low error.

\bibliography{cat_bib}

\end{document}